\documentclass[final,10pt]{elsarticle}

\usepackage{amssymb}
\journal{International Journal of Modern Physics B}

\begin{document}

\begin{frontmatter}

\markboth{N.-C. Yeh et al.}
{Unconventional Low-Energy Excitations of Cuprate Superconductors}

%%%%%%%%%%%%%%%%%%%%% Publisher's Area please ignore %%%%%%%%%%%%%%%
%
%\catchline{}{}{}{}{}
%
%%%%%%%%%%%%%%%%%%%%%%%%%%%%%%%%%%%%%%%%%%%%%%%%%%%%%%%%%%%%%%%%%%%%

\title{UNCONVENTIONAL LOW-ENERGY EXCITATIONS OF\\
CUPRATE SUPERCONDUCTORS}

\author{N.-C. YEH\footnote{Corresponding author. E-mail: ncyeh@caltech.edu}
 and A. D. BEYER}

\address{Department of Physics, California Institute of Technology\\
Pasadena, CA 91125, USA}

%\maketitle

%\begin{history}
%\received{(17 August 2009)}
%\revised{(Day Month Year)}
%\accepted{(?? ?? 2009)}
%\comby{(xxxxxxxxxx)}
%\end{history}

\begin{abstract}
Recent development in the physics of high-temperature cuprate superconductivity is reviewed, with special emphasis on the phenomena of unconventional and non-universal low-energy excitations of hole- and electron-type cuprate superconductors and the possible physical origin. A phenomenology based on coexisting competing orders with cuprate superconductivity in the ground state appears to provide consistent account for a wide range of experimental findings, including the presence (absence) of pseudogaps and Fermi arcs above the superconducting transition $T_c$ in hole-type (electron-type) cuprate superconductors and the novel conductance modulations below $T_c$, particularly in the vortex state. Moreover, the competing order scenario is compatible with the possibility of pre-formed Cooper pairs and significant phase fluctuations in cuprate superconductors. The physical implications of the unified phenomenology and remaining open issues for the microscopic mechanism of cuprate superconductivity are discussed.
\end{abstract}


\end{frontmatter}

\section{Introduction -- Emergent Superconductivity and Competing Orders from Doped Mott Insulators}

Since the discovery of high-temperature superconducting cuprates in 1986~\cite{Bednorz_Muller86}, much progress has been made in the physics and applications of these novel superconductors~\cite{LeePA06,Kivelson03,Sachdev03,Demler04,Damascelli03,Fischer07,Yeh02a}. The complexity of these materials and their strong electronic correlation give rise to various unconventional low-energy excitations and many non-universal physical phenomena amongst hole- and electron-type cuprates~\cite{Yeh05,Yeh07,Yeh09}. These puzzling findings have hindered successful development of a microscopic theory to consistently account for all the differences found in the cuprates of varying doping levels. Therefore, a strategic approach to unravel the mystery of high-temperature superconductivity appears to hinge upon identifying the underlying physical origin for all unconventional and non-universal phenomena among the cuprates. A successful phenomenology that consistently account for all experimental findings will likely provide crucial insights into the microscopic mechanism for cuprate superconductivity.

The objective of this review is to survey the up-to-date status of experimental manifestations of unconventional low-energy excitations in the cuprates, compare two leading theoretical scenarios with empirical findings, explore the feasibility of a unified phenomenology for all cuprates, and discuss the implications of the phenomenology in the context of microscopic pairing mechanism of cuprate superconductiviy. We suggest that the scenario of coexisting competing orders (COs) and superconductivity in the ground state of cuprate superconductors may account for the appearance of unconventional low-energy excitations, novel vortex dynamics and various non-universal phenomena among electron- and hole-type cuprates of different doping levels. Moreover, the occurrence of competing orders does not exclude the possibility of pre-formed Cooper pairs above the superconducting transition as suggested by some other theoretical and experimental evidences. On the other hand, the assumption of a pure superconducting ground state in the cuprates would lead to a number of contradictory findings that cannot be reconciled with experimental observation. We therefore suggest that at low temperatures there is important interplay between the collective low-energy bosonic excitations from COs and Cooper pairs in the superconducting state of the cuprates.

This article is structured as follows. In Section 2 we summarize the most prominent examples of unconventional low-energy excitations in the cuprates as manifested by various experiments. The basic concepts of the one-gap and two-gap phenomenological models are described in Section 3, together with evaluations of their consistency with experimental findings. An overview of latest experimental progress in the studies of scanning tunneling spectroscopy (STS) and angle-resolved photoemission spectroscopy (ARPES) on the low-energy charge excitations of both hole- and electron-type cuprates and a phenomenology that unifies these experimental findings are given in Section 4. In Section 5 important open issues in high-temperature superconductivity and possible clues are discussed. Finally, Section 6 concludes the status of our current understanding of cuprate superconductivity and the challenges required to unravel the mystery of the pairing mechanism.

\section{Unconventional Low-Energy Excitations in the Cuprates}

A salient characteristic associated with all high-temperature superconducting cuprates is that they are doped Mott insulators with strong electronic correlation.~\cite{LeePA06} Mott insulators differ from conventional band insulators in that the latter are dictated by the Pauli exclusion principle when the highest occupied band contains two electrons per unit cell, whereas the former are associated with the existence of strong on-site Coulomb repulsion such that double occupancy of electrons per unit cell is energetically unfavorable and the electronic system behaves like an insulator rather than a good conductor at half filling. An important signature of doped Mott insulators is the strong electronic correlation among the carriers due to poor screening and the sensitivity of their ground state to the doping level. In cuprates, the ground state of the undoped perovskite oxide is an antiferromagnetic Mott insulator, with nearest-neighbor Cu$^{2+}$-Cu$^{2+}$ antiferromagnetic exchange interaction in the CuO$_2$ planes.~\cite{Kastner98} Depending on doping with either electrons or holes into the CuO$_2$ planes,~\cite{Kastner98,Maple90} the N\'eel temperature ($T_N$) for the antiferromagnetic-to-paramagnetic transition decreases with increasing doping level. Upon further doping of carriers, long-range antiferromagnetism vanishes and is replaced by superconductivity. As schematically illustrated in the phase diagrams for the hole- and electron-type cuprates in Fig.~1, the superconducting transition temperature ($T_c$) first increases with increasing doping level ($\delta$), reaching a maximum $T_c$ at an optimal doping level, then decreases and finally vanishes with further increase of doping.

\begin{figure}[!]
\centering
\includegraphics[keepaspectratio=1,width=10cm]{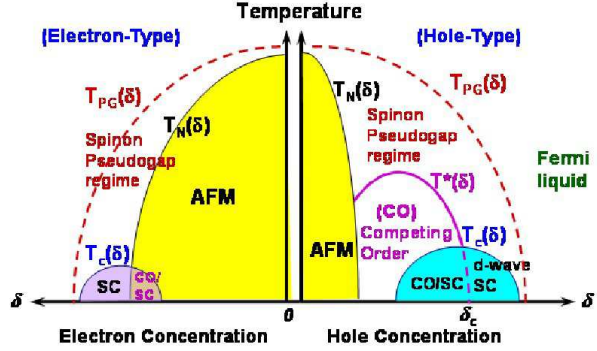}
\caption{A proposed zero-field temperature ($T$) vs. doping level ($\delta$) generic phase diagram for electron- and hole-type cuprates. AFM: antiferromagnetism, CO: competing order, SC: superconductivity, $\delta$: doping level,  $T_N$: N\'eel temperature, $T_c$: superconducting transition temperature, $T^{\ast}$: low-energy pseudogap temperature, $T_{\rm PG}$: high-energy pseudogap temperature.}
\end{figure}

Closer inspection of the phase diagrams for hole- and electron-type cuprates reveals that they are asymmetric: For hole-type cuprates in the under- and optimally doped regime, the physical properties above $T_c$ but below a crossover temperature $T^{\ast}$, known as the low-energy pseudogap (PG) temperature, are significantly different from those of Fermi liquids, including slightly suppressed electronic density of states (DOS) referred to as the low-energy pseudogap phenomenon~\cite{Damascelli03,Fischer07,Norman07} and incompletely recovered Fermi surfaces in the momentum space known as the Fermi arc phenomenon.~\cite{Damascelli03,LeeWS07} Specifically, the PG phenomenon refers to the observation of a soft gap without coherence peaks in the quasiparticle excitation spectra above $T_c$ in hole-type cuprates and below a PG temperature $T^{\ast}$. Evidence for the PG in hole-type cuprates has been reported in tunneling measurements,~\cite{Fischer07,Renner98,Vershinin04,Boyer07,Gomes07} as exemplified in Fig.~2(a) for an underdoped $\rm Bi_2Sr_2CaCu_2O_x$ (Bi-2212) sample, $^{63}$Cu spin-lattice relaxation rate and $^{63}$Cu Knight shift in nuclear magnetic resonance (NMR) experiments,~\cite{Timusk99,Walstedt91,Stern94,Corey96,Julien96,Hsueh97} optical conductivity experiments,~\cite{Timusk99,Puchkov96} Raman scattering experiments~\cite{Opel00,LeTacon06} and angle resolved photoemission spectroscopy (ARPES) measurements.~\cite{Damascelli03,LeeWS07} Intimately related to the PG is the observation of ``Fermi arcs'' in hole-type cuprates, which refers to an incomplete recovery of the full Fermi surface for temperature in the range $T_c < T < T^{\ast}$, as schematically illustrated in Figs.~3(a) -- 3(c). The phenomena is most notably observed in $\rm Bi_2Sr_2CaCu_2O_x$ (Bi-2212) as a function of doping.~\cite{Damascelli03,LeeWS07} The persistence of gapped quasiparticle spectral density functions near the $(\pi, 0)$ and $(0,\pi)$ portions of the Brillouin zone above $T_c$ in hole-type cuprates are the source of the incomplete recovery of the Fermi surface. In contrast, electron-type cuprates exhibit neither the low-energy pseudogap nor the Fermi arc above $T_c$~\cite{Kleefisch01,ChenCT02,Matsui05} (see Figs.~3(d) -- 3(f)), although ``hidden pseudogap'' features in the quasiparticle excitation spectra have been observed under the superconducting dome in doping dependent grain-boundary tunneling experiments on $\rm Pr_{2-x}Ce_xCuO_{4-y}$ (PCCO) and $\rm La_{2-x}Ce_xCuO_{4-y}$ (LCCO) when a magnetic field $H > H_{c2}$ is applied to suppress superconductivity.~\cite{Krockenberger03}

\begin{figure}[!]
\centering
\includegraphics[keepaspectratio=1,width=12cm]{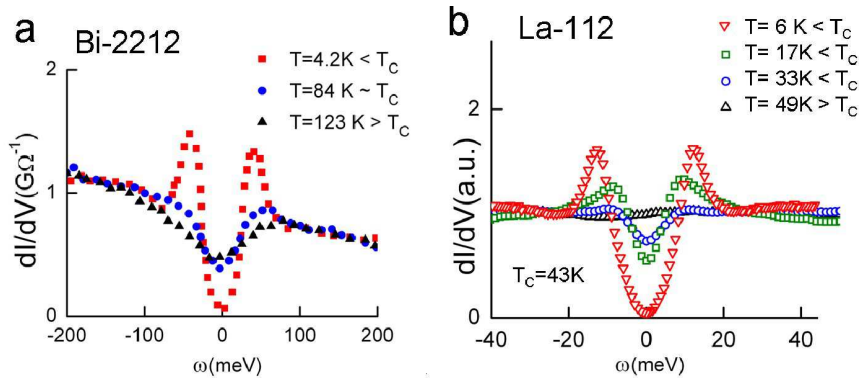}
\caption{Comparison of the temperature dependent scanning tunneling spectra of (a) an underdoped hole-type cuprate Bi-2212 and (b) an optimally doped electron-type cuprate $\rm La_{0.1}Sr_{0.9}CuO_2$ (La-112),~\cite{ChenCT02,Teague09} showing the presence of pseudogap above $T_c$ in the former and the absence of pseudogap above $T_c$ in the latter.}
\end{figure}

\begin{figure}[!]
\centering
\includegraphics[keepaspectratio=1,width=12cm]{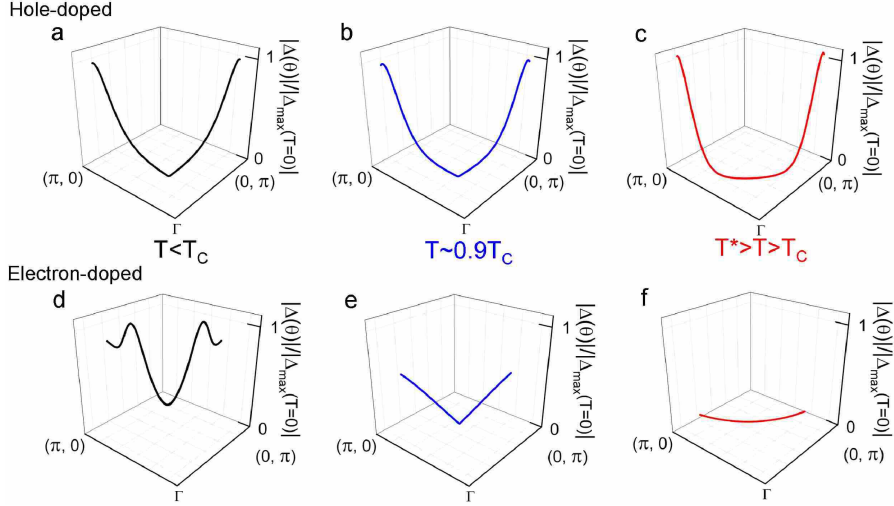}
\caption{Contrasts of the temperature dependent ARPES data between hole- and electron-type cuprate superconductors: (a) -- (c) illustrate the temperature evolution of the momentum dependent effective gap $\Delta _{\rm eff}(\textbf{k})$ determined from ARPES for a hole-type cuprate such as the Bi-2212 system. (d) -- (f) illustrate the temperature evolution of $\Delta _{\rm eff}(\textbf{k})$ in a typical electron-type cuprate. The Fermi arc only exists in hole-type cuprates for $T^{\ast} > T > T_c$.}
\end{figure}

In addition to the low-energy PG phenomenon that is correlated with the Fermi arcs and only found in hole-type cuprates slightly above $T_c$ as shown in Figs.~2(a)-(b) and 3(a)-(f) , there is a high-energy PG that is present in both electron- and hole-type cuprates according to optical~\cite{Onos01,Gallais05} and neutron scattering~\cite{Yamada98,Motoyama06} experiments. As elaborated further in this review, the low-energy PG seems to be associated with the onset of competing orders while simultaneously accompanied by strong phase fluctuations. In contrast, the high-energy pseudogap appears to be related to the short-range magnetic exchange coupling in the cuprates.

In addition to the contrasting low-energy charge excitations among the hole- and electron-type cuprates, the low-energy spin excitations are also different: Neutron scattering experiments on the hole-type $\rm La_{2-x}Sr_xCuO_{4-y}$ reveal incommensurate spin correlations in the superconducting state, with a temperature independent spin gap observed both below and above $T_c$.~\cite{Yamada95,Lake99,LeeCH00} In contrast, the one-layer electron-type cuprate $\rm Nd_{2-x}Ce_xCuO_{4-y}$ (NCCO)~\cite{Yamada03} displayed commensurate spin correlations and a temperature dependent spin gap in neutron scattering experiments.~\cite{Yamada03} Furthermore, the spin gap of NCCO was observed to be a maximum of about 4 meV as $T \to 0$ and disappeared as $T \to T_c$.~\cite{Yamada03} Therefore, the spin gap in both hole- and electron-type cuprates may possibly be related to the pseudogap, as both the spin gap and pseudogap are absent above $T_c$ in electron-doped cuprates and present above $T_c$ in hole-doped cuprates. Moreover, enhancement of a static commensurate magnetic order by application of a magnetic field up to 9 T in electron-doped $\rm Pr_{0.89}LaCe_{0.11}CuO_4$ (PLCCO) is observed,~\cite{Fujita04} and a commensurate quasi-2D spin-density wave enhanced by the application of a magnetic field equal to 5 T in underdoped $\rm Pr_{0.88}LaCe_{0.12}CuO_4$ is also reported.~\cite{Kang05} Neither the orders observed by neutron scattering nor the pseudogap are suppressed by moderate magnetic fields. However, these results are not conclusive for relating the pseudogap to the spin gap because it has been cautioned that secondary material phases associated with paramagnetism of Nd$^{3+}$ or Pr$^{3+}$ could form during oxygen reduction in the synthesis of one-layer electron-doped cuprates. Thus, whether holes are the actual carriers in one-layer electron-doped cuprates must be further examined.~\cite{Mang04,Richard04} Fortunately, these issues do not arise in the infinite-layer electron-type cuprate $\rm La_{0.1}Sr_{0.9}CuO_2$ (La-112) system because oxygen reduction is not required to achieve optimal doping,~\cite{Jung02a} La-112 is free from magnetic ions, and thermopower measurements have verified that electrons are the relevant carriers in La-112.~\cite{Williams02}

While the exact ground state and the corresponding low-energy excitations of a given cuprate is determined by such material parameters as the level of electron or hole doping, the electronic anisotropy, the number of CuO$_2$ layers per unit cell and the detailed bandstructures of the material, most cuprate superconductors differ fundamentally from conventional superconductors in that their ground states are not comprised of a pure superconducting phase. Various ``competing orders'' (COs) have been found to coexist with superconductivity (SC) in the ground state, as manifested by such experimental evidences as scanning tunneling spectroscopy (STS),~\cite{Boyer07,Hoffman02,Howald03,Hanaguri04,Wise08,Beyer09,Teague09} neutron scattering,~\cite{Motoyama06,Tranquada95,Tranquada97,Wells97,Lake01,Mook02,Fujita02,Matsuda02} muon spin resonance ($\mu$SR),~\cite{Tokiwa03} NMR,~\cite{Kotegawa01,Kotegawa01a} optical and Raman scattering measurements~\cite{Opel00,LeTacon06} and ARPES.~\cite{LeeWS07,Kondo07} The existence of competing orders has been further confirmed by theoretical modeling and simulations.~\cite{Kivelson03,Sachdev03,Demler04,Zhang97,Vojta00,Varma97,Demler01,Polkovnikov02,ChenHD02,ChenHD04,Chakravarty01,Schollwock03,LiJX06} A commonality among these experimental findings is that there appears to be an energy scale and/or a collective mode associated with an order parameter that exhibits temperature, magnetic field and/or momentum dependences different from those of the superconducting order parameter. The occurrence of specific types of collective modes such as the charge-density waves (CDW),~\cite{Kivelson03,LiJX06} pair-density waves (PDW),~\cite{ChenHD02,ChenHD04} or spin-density waves (SDW)~\cite{Demler04,Demler01,Polkovnikov02} depends on the microscopic properties of a given cuprate, and is therefore non-universal.~\cite{Yeh05,Yeh07} In particular, a number of contrasting properties between hole- and electron-type cuprates, as exemplified by the temperature ($T$) vs. doping level ($\delta$) phase diagrams in Fig.~1, cannot be explained by a ground state of superconductivity alone. In addition to the apparent disparity of the highest superconducting transition temperatures, the most noticeable contrasting properties include the following: the presence or absence of the low-energy PG,~\cite{Damascelli03,Fischer07,ChenCT02,ChenCT07,Beyer08} anomalous Nernst effect~\cite{WangY06} and Fermi arcs~;\cite{Damascelli03,LeeWS07,Matsui05,YuBL09} varying degrees of spatial homogeneity and modulations in the quasiparticle spectra;~\cite{Beyer09,Teague09,Beyer08,Yeh01,Bobroff02,McElroy05,Lang02} and the characteristics of magnetic excitations.~\cite{Motoyama06,DaiP01,Norman00} Therefore, a successful microscopic theory must account for all these non-universal phenomena and unconventional low-energy excitations among the cuprates in addition to addressing the mechanism of Cooper pairing.

Although the relevance of competing orders to the occurrence of cuprate superconductivity remains unclear to date, the existence of competing orders has a number of important physical consequences. In addition to the aforementioned non-universal phenomena among different cuprates, quantum criticality naturally emerges as the result of competing phases in the ground state.~\cite{Sachdev03,Demler04,Yeh05,Varma97}. Moreover, strong quantum fluctuations are expected due to proximity to quantum criticality~\cite{Yeh05,Yeh07,Beyer07,Zapf05} Macroscopically, the presence of COs and strong quantum fluctuations naturally lead to weakened superconducting stiffness upon increasing $T$ and magnetic field $H$,~\cite{Yeh05,Beyer07,Emery95,Corson99} which may be responsible for the extreme type-II nature and the novel vortex dynamics of cuprate superconductors,~\cite{Beyer07,Zapf05,Blatter94,FisherDS91,Balents94,Balents95,Giamarchi94,Giamarchi95,Yeh93} as schematically illustrated in Figs.~4(a) and 4(b) for the magnetic field ($H$) vs. temperature ($T$) vortex phase diagrams of cuprate superconductors for $H \parallel \hat c$-axis and $H \perp \hat c$-axis, respectively. Additionally, the low-energy excitations from the ground state become unconventional due to the redistributions of the spectral weight between SC and COs in the ground state.~\cite{Yeh07,Beyer09,Teague09,ChenCT07,Beyer08} Among the best known phenomena associated with unconventional low-energy excitations include: the appearance of satellite features~\cite{Fischer07,Yeh05,Yeh07,ChenCT07,Beyer08} and periodic LDOS modulations in the quasiparticle spectra below $T_c$~\cite{Boyer07,Hoffman02,Howald03,Hanaguri04,Wise08,Beyer09,Teague09} and the related existence of a low-energy PG and Fermi arcs for $T_c < T < T^{\ast}$~\cite{ChenCT07,YuBL09} in hole-type cuprates; excess sub-gap quasiparticle density of states (DOS) in electron-type cuprates below $T_c$;~\cite{Yeh05,Yeh07} ``dichotomy'' in the momentum dependence of quasiparticle coherence in hole-type cuprates;~\cite{Damascelli03,ChenCT07,McElroy05,ZhouXJ04} and PG-like vortex-core states in both electron- and hole-type cuprate superconductors.~\cite{Beyer08,Beyer09,Teague09,Pan00} Numerous attempts have been made to explain the anomalous behavior amongst the cuprates, and most theoretical models to date may be categorized into two types of scenarios, which we refer to as the ``one-gap'' and ``two-gap'' models. The former is associated with the ``pre-formed pair'' conjecture that asserts strong phase fluctuations in the cuprates so that formation of Cooper pairs may occur at a temperature well above the superconducting transition. The latter considers coexistence of CO and SC in the ground state of the cuprates.

\begin{figure}[!]
\centering
\includegraphics[keepaspectratio=1,width=11cm]{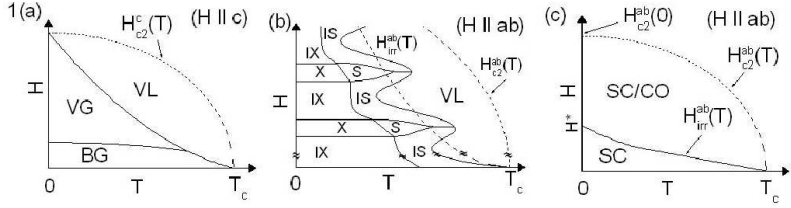}
\caption{Schematic $H$-vs.-$T$ vortex phase diagrams of cuprate superconductors: (a) For $\textbf{H} \parallel \hat c$, assuming thermal fluctuations and relevant random point disorder.~\cite{Blatter94,FisherDS91,Giamarchi94,Giamarchi95} The vortex phase (neglecting the lower critical field $H_{c1}$ due to the extreme type-II nature) with increasing $H$ and $T$ evolves from the Bragg glass (BG) to the vortex glass (VG) and then the vortex liquid (VL) before reaching the upper critical field $H_{c2} ^c$. (b) Vortex phase diagram for $\textbf{H} \parallel ab$ (or equivalently, $\textbf{H} \perp \hat c$), assuming dominating thermal fluctuations and intrinsic pinning effects due to the periodic CuO$_2$ planes while neglecting quantum fluctuations.~\cite{Balents94,Balents95} (S: commensurate vortex smectic phase, IS: incommensurate vortex smectic phase, VL: vortex liquid, X: commensurate vortex crystal, IX: incommensurate vortex crystal). The dashed curve represents the phase boundary modified by magnetic field misalignment and stacking faults that tend to smear the modulations of the S/IS to VL phase boundary and reduce the phase boundary into being monotonically dependent on $H$. (c) For $\textbf{H} \parallel ab$, assuming dominating quantum fluctuations associated with the proximity to quantum criticality and competing orders.~\cite{Yeh05,Yeh07,Beyer07,Zapf05}}
\end{figure}

\section{``One-Gap'' versus ``Two-Gap'' Models}

Among the unconventional and non-universal cuprate phenomena, the PG and Fermi arc phenomena as well as the spectral inhomogeneity observed in some hole-type cuprates have attracted much of the theoretical attention. In particular, the occurrence of the PG above $T_c$ in the underdoped hole-type cuprates has been considered by the one-gap model as the necessary precursor of cuprate superconductivity~\cite{LeePA06,Norman07,Emery95,Gros06,Chubukov07}. In this scenario, cuprate superconductivity evolves smoothly from a high-temperature normal state to an intermediate-temperature pre-formed pair regime with strong phase fluctuations, and finally to a low-temperature superconducting regime with condensed pairs and global phase coherence. Therefore, the PG temperature $T^{\ast}$ is attributed to the pair formation temperature in this scenario and the superconducting temperature $T_c$ to the pair condensation temperature. However, the increasing $T^{\ast}$ with decreasing doping level anti-correlates with the doping dependence of $T_c$ in hole-type cuprates. Moreover, the absence of $T^{\ast}$ in all electron-type cuprates clearly contradicts the notion of pre-formed pairs being a necessary condition for cuprate superconductivity. Above all, the one-gap model cannot provide natural explanations for all of the non-universal and unconventional low-energy excitations (e.g. the appearance of energy-independent wave-vectors or charge modulations in the SC state of hole-type cuprates~\cite{Boyer07,Hoffman02,Wise08,Beyer09}) observed among different cuprates.

Another school of thoughts may be referred to as the two-gap (or competing-order) model, which attributes the low-energy PG phenomena to the occurrence of CO phase instabilities.~\cite{Kivelson03,Demler04,Yeh05,Yeh07,Zhang97,Varma97,Demler01,Polkovnikov02,ChenY02,ChenHD02,ChenHD04,Chakravarty01,LiJX06} In this scenario, the unconventional physical properties observed for $T_c < T < T^{\ast}$ are associated with the onset of CO so that SC and CO become coexist below $T_c$ and in the ground state. While the relevance of CO to SC remains unclear, the CO scenario is not exclusive of the possibility of preformed pairs:~\cite{Yeh09,Beyer09} competing orders represent additional phase instabilities in the cuprates that are neglected in all one-gap models, and they may naturally coexist with pre-formed pairs above $T_c$ if they can already coexist with condensed Cooper pairs at low temperatures. Additionally, various unconventional low-energy excitations and non-universal phenomena among different cuprates may be consistently accounted for in the context of CO scenario.~\cite{Yeh09,Beyer09,Teague09,ChenCT07,Beyer08,YuBL09}

While phenomenologically successful, the greatest difficulty encountered in the two-gap model is to find a microscopic mechanism that correctly predicts the occurrence of a specific type of CO phase instability for a given set of materials parameters and simultaneously accounts for the interplay of CO and SC at varying temperatures and magnetic fields. In the following subsections, we examine the rationale and basic concepts of the one-gap and two-gap models, and then point out the difficulties and open issues associated with both schools of thoughts.

\subsection{Basic concepts and open issues of the one-gap scenario}

The concept of the one-gap scenario may be traced back to Anderson's idea of the resonating valence bond (RVB) for pairing of singlet spins in a square lattice.~\cite{LeePA06,AndersonPW87,AndersonPW87a,Affleck88} The RVB state is a highly degenerate and strongly fluctuating spin-liquid phase that is most applicable to the cuprates in the insulating limit. It exhibits spin-1/2 fermionic excitations known as spinons that carry no charge. In reality, however, the RVB state does not account for the Mott insulating limit of the cuprates correctly. As shown in Fig.~1, neutron scattering experiments have clearly demonstrated that the insulating limit of both the electron- and hole-type cuprates is a Mott antiferromagnetic (AFM) insulator with ordered spins rather than strongly fluctuating spin liquids. Subsequent improvements of the RVB scenario employ a mean-field slave-boson technique~\cite{Baskaran87} and a projected wave-function method~\cite{Gros88,Gros89} with analytic and numerical treatments of the t-J model to enforce the constraint againt double occupation in the Mott insulating limit and to put the RVB idea on a more formal footing. The implementation of this local constraint is shown to naturally lead to gauge theories.~\cite{LeePA06} Further inclusion of gauge fluctuations are argued to provide qualitative accounts for the phase fluctuations in the pseudogap regime of the cuprates,~\cite{Baskaran88,Ioffe89,Nagaosa90} although it is also noted that phase fluctuations can only explain pseudogap phenomenology over a limited temperature range, and some additional physics is needed to explain the onset of singlet formation at much higher temperatures in the strongly underdoped limit.~\cite{LeePA06} Additionally, the introduction of gauge theory naturally differentiates between the high-energy gauge group in the formulation of the problem versus the low-energy gauge group that is an emergent phenomenon. Therefore, deconfinement based on different emergent gauge groups would occur at low temperatures, leading to fractionalization and spin-charge separation.~\cite{LeePA06} Moreover, superconductivity may emerge from the hole-doped RVB state with $d_{x^2-y^2}$-wave symmetry and a state with spin-gap properties above $T_c$ in the underdoped region.~\cite{Kotliar88}

Theoretically, a limitation to the rigorous formalism of the one-gap model based on the concept of a deconfined spin liquid is the strict no-double occupancy constraint only applicable to the insulating limit. Hence, it seems difficult to generalize the physics derived from the insulating limit to cuprate superconductors that are of significant doping of carriers so that the no-double occupancy constraint becomes relaxed. Empirically, the important signature of fractionalization and spin-charge separation associated with the one-gap model has never been realized. Moreover, small zero-field quantum fluctuations in the ground state~\cite{McElroy05,Lang02} and strong magnetic field-induced quantum fluctuations of most cuprate superconductors~\cite{Beyer07,Zapf05} are inconsistent with the notion of a spin liquid. Above all, the necessary condition of a pseudogap phase and strong phase fluctuations asserted by the pre-formed pair scenario is completely inapplicable to electron-type cuprate superconductors.~\cite{Kleefisch01,ChenCT02,Matsui05,Krockenberger03} Unless one is willing to settle with different pairing mechanisms for electron- and hole-type cuprate superconductors, these disagreements between empirical facts and theoretical predictions of the one-gap model certainly call for in-depth reevaluations of the basic theoretical assumption.

In addition to the aforementioned efforts that attempt to establish a rigorous microscopic theoretical foundation for the one-gap model, there are different approaches based on either phenomenological coupling between spins and fermions or via the random-phase approximation treatment for the Hubbard model, which is effectively a weak-coupling expansion. These approaches are explored for the possibility of $d_{x^2-y^2}$-wave superconductivity in terms of the exchange of spin fluctuations.~\cite{Scalapino95} Other phenomenological models have also been proposed to account for different experimental findings.~\cite{Norman07,McElroy05,Chubukov07,McElroy03,Alldredge08} Unfortunately, these phenomenological models with ad hoc assumptions are only applicable to explaining a specific set of experimental results within a limited parameter space, and they generally become incompatible with global comparison of experimental results as functions of the doping level, number of CuO$_2$ layers per unit cell, temperature, and magnetic field. In the following, examples of inconsistency among different phenomenolofical one-gap models are discussed.

As the first example, the one-gap model has asserted that quasiparticles are well defined below $T_c$ so that all low-energy excitations may be understood in terms of properties of Bogoliubov quasiparticles associated with pure $d_{x^2-y^2}$-wave superconductivity.~\cite{Norman07,McElroy05,Chubukov07,Alldredge08} In this context, the Fourier-transformed (FT) local density of states (LDOS) of Bi-2212 systems obtained from STS studies at $T \ll T_c$ have been interpretted as manifestations of elastic quasiparticle scattering interferences associated with eight energy-dependent impurity-scattering wave-vectors that connect the highest joint density of states,~\cite{McElroy05} as illustrated in Figs.~5(a) and 5(c). This analysis of STS data is also known as the ``octet model''.

\begin{figure}[!]
\centering
\includegraphics[keepaspectratio=1,width=10cm]{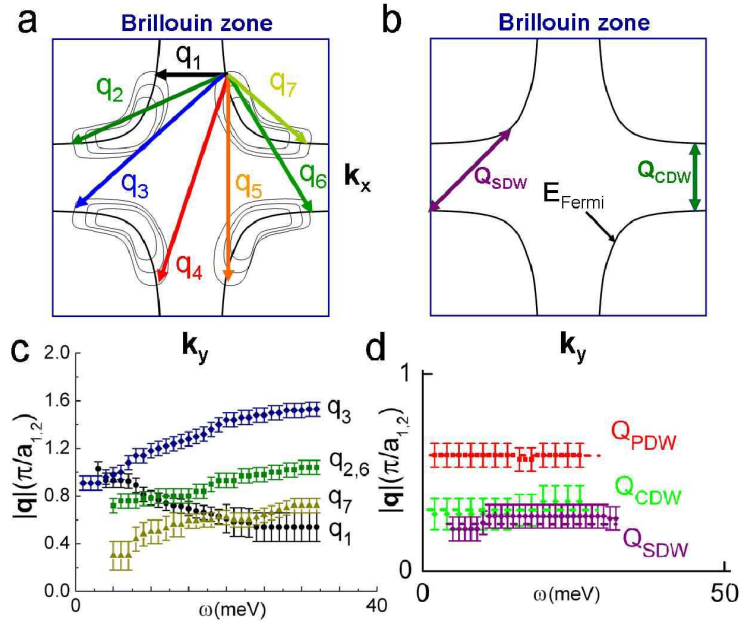}
\caption{Modes of quasiparticle scattering interferences and collective excitations: (a) Illustration of the wave-vectors associated with elastic quasiparticle scattering interferences between pairs of points on equal energy contours with maximum joint density of states. (b) Illustration of the wave-vectors associated with SDW and CDW excitations. (c) The momentum ($|\textbf{q}_i|$) vs. energy ($\omega$) dispersion relations of the quasiparticle scattering wave-vectors derived from the FT-LDOS data of Y-123.~\cite{Beyer09} (d) The nearly energy-independent collective modes $|\textbf{Q}_{\rm PDW}|$, $|\textbf{Q}_{\rm CDW}|$ and $|\textbf{Q}_{\rm SDW}|$ obtained from the FT-LDOS data of Y-123.~\cite{Beyer09}}
\end{figure}

Although this octet model enjoys certain degrees of success in accounting for the momentum ($|\textbf{q}|$) vs. energy ($\omega$) dispersion relation of some of the modes for Bi-2212 systems~\cite{McElroy05,McElroy03} and also examplified in Fig.~5(c) for an optimally doped Y-123,~\cite{Beyer09} several irreconcilable difficulties also arise. First, one of the modes (denoted as $\textbf{q}_1$) along the Cu-O bonding direction should not have been present had the ground-state of Bi-2212 been purely superconducting, because the superconducting coherence factor associated with Bogoliubov quasiparticles would have extinguish the intensity of the $\textbf{q}_1$ mode.~\cite{ChenCT03} The experimental findings of nearly energy-independent $q_1$ mode either implies the sample being non-superconducting or other mechanisms being present to result in strong intensities associated with the $\textbf{q}_1$ mode.~\cite{ChenCT03} Second, further detailed evaluations of the FT-LDOS data revealed that the octet model of simple quasiparticle scattering interferences in fact only worked well for quasiparticle momenta closer to the nodal direction; experimental data began to deviate significantly for quasiparticle momenta closer aligned along the Cu-O bonding directions.~\cite{Hanaguri09} Third, apparent existence of ``checkerboard-like'' LDOS modulations has been found both in zero and finite magnetic fields,~\cite{Vershinin04,Boyer07,Hoffman02,Hanaguri04,Wise08,Beyer09} with better enhanced LDOS modulations within the vortex cores in finite fields.~\cite{Hoffman02,Beyer09} In Fig.~6 exemplified LDOS modulations obtained in both $H = 0$ and $H = 5$ T for optically doped Y-123 are shown.~\cite{Yeh09,Beyer09} Furthermore, the wave-vector of the LDOS modulations is found to be doping dependent and nested to the Fermi surface in the $\rm Bi_2Sr_2CuO_x$ (Bi-2201) system,~\cite{Boyer07,Wise08} whereas in the Bi-2212 system the modulations even survived above $T_c$ in the pseudogap regime.~\cite{Vershinin04} These novel low-energy excitations cannot be easily reconciled with the behavior of simple Bogoliubov quasiparticle excitations from a pure superconducting ground state.

\begin{figure}[!]
\centering
\includegraphics[keepaspectratio=1,width=10cm]{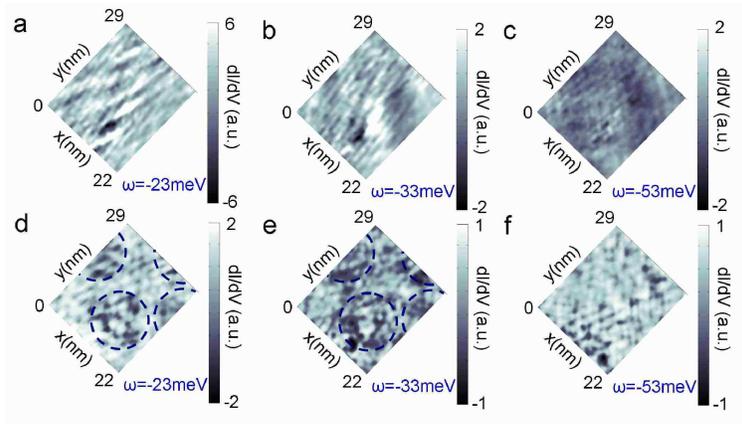}
\caption{The LDOS modulations of Y-123 for $H$ = 0 and 5 Tesla and $T = 6$ K over a $(22 \times 29) \rm nm^2$ area:~\cite{Yeh09,Beyer09} Patterns associated with density-wave modulations for $H$ = 0 at $\omega =$ (a) $\rm - 23 meV \sim -\Delta _{\rm SC}$, (b) $\rm - 33 meV \sim - V_{\rm CO}$, and (c) $\rm - 53 meV$. In contrast, at a finite magnetic field $H$ = 5 Tesla, vortices (circled objects) in addition to density-wave modulations are visible, as exemplified in (d) for $\omega = - 23$ meV, (e) for $-33$ meV , and (f) for $-53$ meV. We note that the vortex contrasts are the most apparent at $|\omega| <~ \Delta _{\rm SC}$ and become nearly invisible for $|\omega| \gg V_{\rm CO}$. The vanishing contrast at high energies may be attributed to the onset of Cu-O optical phonons ($\sim 50$ meV for the cuprates~\cite{LeeJ06}) so that both the collective modes and quasiparticles become scattered inelastically.}
\end{figure}

The second example is related to the one-gap Fermi arc model that employs the Green function techniques.\cite{Norman07} To begin, the low-energy effective superconducting Hamiltonian ${\cal H}_{\rm SC}$ for a pure superconductor may be written as
\begin{eqnarray}
{\cal H}_{\rm SC} &= \sum _{{\rm k},\sigma} \xi _{\rm k} c^{\dagger} _{{\rm k},\sigma} c_{{\rm k},\sigma}
- \sum _{{\rm k}} \Delta _{\rm SC} (\textbf{k}) (c^{\dagger} _{{\rm k},\uparrow}  c^{\dagger}_{-{\rm k},\downarrow} + c_{-{\rm k},\downarrow} c_{{\rm k},\uparrow}), \nonumber\\
 &= \sum _{{\rm k}} \left( c^{\dagger} _{{\rm k},\uparrow} c_{-{\rm k},\downarrow} \right)
\pmatrix{\xi _{{\rm k}} &-\Delta _{\rm SC} (\textbf{k})\cr
              -\Delta _{\rm SC} (\textbf{k}) &-\xi _{\rm k}\cr}
      \pmatrix{c_{{\rm k},\uparrow}\cr
                    c^{\dagger}_{-{\rm k},\downarrow}\cr},
\label{eq:HSC}
\end{eqnarray}
where $\xi _{\rm k}$ is the normal-state eigen-energy for particles of momentum $\textbf{k}$ relative to the Fermi energy, $\sigma = \uparrow , \downarrow$ refers to the spin states, $c^{\dagger}$ and $c$ are the fermion creation and annihilation operators, and $\Delta _{\rm SC} (\textbf{k})$ denotes the superconducting energy gap. Hence, ${\cal H}_{\rm SC}$ is a $(2 \times 2)$ matrix, and the adjoint of $\Psi _{\rm k}$	represents a $(1 \times 2)$ matrix 	$\Psi ^{\dagger} _{\rm k} \equiv (c^{\dagger} _{{\rm k},\sigma} \ c_{{\rm k},\sigma})$.	The mean-field superconducting Hamiltonian in Eq.~(\ref{eq:HSC}) can be exactly diagonalized so that the Green function for quasiparticles $G_0(\textbf{k},\omega)$ is given by
\begin{equation}
G_0 ^{-1}(\textbf{k},\omega) = \omega I - {\cal H}_{\rm SC}
= \pmatrix{\omega - \xi _{\rm k} &\Delta _{\rm SC} (\textbf{k})\cr
                \Delta _{\rm SC} (\textbf{k}) &\omega + \xi _{\rm k}\cr},
\label{eq:G01}
\end{equation}
where $I$ in Eq.~(\ref{eq:G01}) denotes the $(2 \times 2)$ unit matrix. Given the Green function of a Hamiltonian, the spectral density function $A(\textbf{k},\omega)$ may be obtained according to the relation
\begin{equation}
A(\textbf{k},\omega) \equiv - {\rm Im} \lbrack G_0(\textbf{k},\omega) / \pi \rbrack,
\label{eq:Akw}
\end{equation}
and the quasiparticle density of states ${\cal N} (\omega)$ becomes
\begin{equation}
{\cal N} (\omega) \equiv \sum _{\rm k} A(\textbf{k},\omega).
\label{eq:Nw}
\end{equation}
If one further incorporates a finite quasiparticle linewidth $\Gamma _{\rm k}$,~\cite{Norman07} the value of the Green function may be rewritten into the following form:
\begin{equation}
G_0 ^{-1} (\textbf{k},\omega) = \omega - \xi _{\rm k} + i \Gamma _{\rm k} - \frac{\Delta ^2 _{\rm SC} (\textbf{k})}{\omega + \xi _{\rm k} + i \Gamma _{\rm k}}.
\label{eq:G02}
\end{equation}
Using Eqs.~(\ref{eq:G02}) and (\ref{eq:Akw}), the spectral density function becomes~\cite{Norman07}
\begin{equation}
A(\textbf{k},\omega) = - \frac{1}{\pi} \frac{\Gamma _{\rm k} \left( \omega ^2 - \xi ^2 _{\rm k} - \Delta ^2 _{\rm SC} (\textbf{k}) - \Gamma ^2 _{\rm k} \right) - 2 \Gamma _{\rm k} \omega (\omega + \xi _{\rm k} )}{ \left( \omega ^2 - \xi ^2 _{\rm k} - \Delta ^2 _{\rm SC} (\textbf{k}) - \Gamma ^2 _{\rm k} \right) ^2 + 4 \Gamma ^2 _{\rm k} \omega ^2}.
\label{eq:Akw2}
\end{equation}

In Ref.~\cite{Norman07}, Eq.~(\ref{eq:Akw2}) is utilized to model the Fermi arc phenomena in the cuprates. In contrast to the expectation of a conventional superconductor, the value of $\Gamma _{\rm k}$ is assumed to be finite and constant as a function of $\textbf{k}$. Thus, the resulting spectral density functions based on the assumption of finite $\Gamma _{\rm k}$ for $\xi _{\rm k} = 0$ (i.e., at the Fermi surface) using Eq.~(\ref{eq:Akw2}) yields a Fermi arc-like behavior. On the other hand, for $\Gamma _{\rm k} \to 0$ the Fermi surface would exhibit four nodal points as long as the gap is finite. Hence, a finite $\Gamma _{\rm k}$ in the pseudogap phase is essential to the occurrence of the Fermi arc. However, a natural question as to why Fermi arcs are not observed for $T < T_c$ arises from this analysis. Namely, the Fermi arc only disappears for $\Gamma _{\rm k} = 0$, and a finite Fermi arc will always be present for any finite values of $\Gamma _{\rm k}$. Therefore, to explain the absence of the pseudogap for $T < T_c$, one must assume that $\Gamma _{\rm k}$ immediately drops to zero upon crossing below the superconducting transition temperature. The validity of this assumption is questionable when comparison with empirical facts is made. In particular, for the Bi-2212 system on which the Fermi arc phenomena were based, strong spatial inhomogeneity was found in the tunneling spectra at $T \ll T_c$.\cite{McElroy05,Lang02} This finding implies significantly large values of $\Gamma _{\rm k}$, which is irreconcilable with the assumption of $\Gamma _{\rm k} = 0$ immediately below $T_c$.

As the third example, it is known that the LDOS exhibits strong spatial variations in Bi-2212 systems, with the degree of spectral inhomogeneity increasing with decreasing doping.\cite{McElroy05,Lang02} In nominally overdoped systems, the tunneling spectra are mostly homogeneous and exhibit a set of sharp superconducting coherent peaks. In optimally doped systems, most of the tunneling spectra exhibit combinations of a set of sharp superconducting coherent peaks and additional ``hump''-like satellite features at higher energies, with the sharp features relatively homogeneous and the satellite features more inhomogeneous. With further decreasing doping, the ``hump''-like features mostly evolve into large and broad peaks, which either render the sharp SC coherence peaks into small ``kinks'' or often overwhelm them. An attempt to account for such strongly spatially varying LDOS in Bi-2212 systems with the one-gap scenario first assumes only one energy gap $\Delta _1$ and one quasiparticle linewidth $\Gamma$ associated with the rounded peak features in underdoped Bi-2212. However, by forcing the one-gap fitting, Ref.~\cite{Alldredge08} eventually finds the necessity of using two effective energy scales to fully account for the experimental data. Speficially, the original functional form of the quasiparticle density of states for pure superconductors with a finite quasiparticle lifetime may be written as:
\begin{equation}
{\cal N} (\omega , \Gamma _{\rm k}) = A \times {\rm Re} \left( \frac{\omega + i \Gamma _1}{\sqrt{(\omega + \Gamma _1)^2} - \Delta ^2 _{\rm SC} (\textbf{k})} \right),
\label{eq:Nw1}
\end{equation}
where $\Gamma _1$ is a constant linewidth term and is equivalent to $\Gamma _{\rm k}$ for the analysis of Ref.~\cite{Norman07} discussed above. Next, the phenomenological model in Ref.~\cite{Alldredge08} assumes that Eq.~(\ref{eq:Nw1}) may be extended to the following ad hoc form
\begin{equation}
{\cal N} (\omega , \Gamma _{\rm k}) = A \times {\rm Re} \left( \frac{\omega + i \Gamma (\omega)}{\sqrt{\left( \omega + \Gamma (\omega) \right)^2} - \Delta ^2 _{\rm SC} (\textbf{k})} + B \times E \right).
\label{eq:Nw2}
\end{equation}
The extension of the functional form in Eq.~(\ref{eq:Nw2}) introduces the term $B \times E$ to mimic the strong particle-hole asymmetry observed in the quasiparticle density of states of Bi-2212~\cite{McElroy05,Lang02,Alldredge08}. Further, it assumes without specific justifications that $\Gamma (\omega)$ is a linear function of energy so that $\Gamma (\omega) \propto \omega$.~\cite{Alldredge08} This ad hoc assumption of linear energy dependence of $\Gamma (\omega)$ is crucial to the analysis of the quasiparticle spectra in Bi-2212 because the zero energy ($\omega = 0$) density of states observed experimentally is equal to zero within experimental uncertainty for all spatially inhomogeneous spectra.~\cite{McElroy05,Lang02} Based on this assumption, the highly broadened peaks, at $|\omega| = \Delta _1$ for quasiparticle spectra with large peak-to-peak gap values, $\Delta _{\rm pk-pk} = \Delta _1$, may be accounted for in addition to the vanishing density of states at $\omega = 0$.

An important feature that the above analysis cannot account for is the homogeneous ``kink''-like feature occurring at energies smaller than the large ``hump'' energy in the quasiparticle spectra in underdoped Bi-2212.\cite{McElroy05} To investigate the kink-like feature further, Ref.~\cite{Alldredge08} computed the average kink-like feature values $\langle \Delta _0 \rangle$ as a function of position from spatially resolved maps and found that the feature was correlated with the superconducting ``dome'' $T_c (\delta)$ shown in Fig.~1. On the other hand, both $\Gamma$ and $\Delta _1$ increase with decreasing doping level, similar to the doping dependence of the low-energy pseudogap. These findings effectively indicate that the one-gap model does not fit any better than a two-gap model, because it is unclear if the behavior of $\Gamma$ is due to disordered superconductivity or competing orders interacting with superconductivity. Further, the fact that $\langle \Delta _0 \rangle$ follows the doping dependence of $T_c$ implies that this feature unaccounted for by the one-gap model is highly relevant to superconductivity. Therefore, an equally valid assumption would have been that $\langle \Delta _1 \rangle$ is either due to a competing order or due to combined effects of superconductivity and a competing order in a ``two-gap'' model, while the kink feature $\langle \Delta _0 \rangle$ neglected in this phenomenological one-gap model~\cite{Alldredge08} actually relates directly to the superconducting gap. Overall, the analysis of Ref.~\cite{Alldredge08} in fact also finds two energy scales in the process of forcing the validity of the one-gap model, which is apparently self contradictory. Additionally, the assumption of a scattering rate $\Gamma \propto \omega$ implies that nodal quasiparticles acquire an infinite lifetime, which contradicts the significant scattering interferences observed for nodal quasiparticles and the octet model prescribed by some of the same authors.~\cite{McElroy05} Moreover, the energy linewidth derived from the forced one-gap fitting generally yields unreasonably large $\Gamma$ at $T \ll T_c$, which would have rendered the SC ground state so disordered that momentum could no longer be a good quantum number, again contradicting the basic assumptions of the octet model. Furthermore, this finding cannot be reconciled with the aforementioned one-gap model for the Fermi arc phenomena~\cite{Norman07} at $T_c < T < T^{\ast}$: According to the assumptions in Ref.~\cite{Norman07}, Fermi arcs vanish at and below $T_c$ due to divergent quasiparticle lifetimes immediately below $T_c$ while the lifetimes were assumed to be finite and independent of momentum above $T_c$. These assumptions are apparently in stark contrast to the results derived from the forced one-gap fitting to the STS data of Bi-2212~\cite{Alldredge08} well below $T_c$.

To quantitatively illustrate the contraditions between the aforementioned Fermi-arc and STS models that are both based on the one-gap scenario, we show in Fig.~7 the Fermi arc behavior predicted by the model for the ARPES data~\cite{Norman07} with different scattering rates. Unless the ad hoc condition for the scattering rate $\Gamma \propto \omega$ is strictly enforced immediately below $T_c$, any finite scattering rate for nodal quasiparticles below $T_c$ would have resulted in finite Fermi arcs, which apparently disagreed with all ARPES experimental findings. That is, to explain the absence of the Fermi arc for $T < T_c$, it must be assumed that $\Gamma _{\rm k}$ in Eq.~(\ref{eq:Akw2}) goes to zero upon crossing below $T_c$ and rapidly increases to a finite and momentum-independent value for $T_c < T < T^{\ast}$. The validity of this assumption is questionable when comparison with empirical facts is made. In particular, for the Bi-2212 system on which the Fermi arc phenomena were based, very large $\Gamma$ values were derived from the one-gap fitting to the tunneling spectra at $T \ll T_c$. This finding would imply a significant value of $\Gamma _{\rm k}$ in Eq.~(\ref{eq:Akw2}) at $T \ll T_c$, which is inconsistent with the assumption of $\Gamma _{\rm k} = 0$ immediately below $T_c$. Finally, neither the one-gap model for ARPES nor the one-gap model for STS data applies to experimental findings in electron-type cuprates, where Fermi arcs and PG are both absent above $T_c$ as shown in Fig.~5.

\begin{figure}[!]
\centering
\includegraphics[keepaspectratio=1,width=8cm]{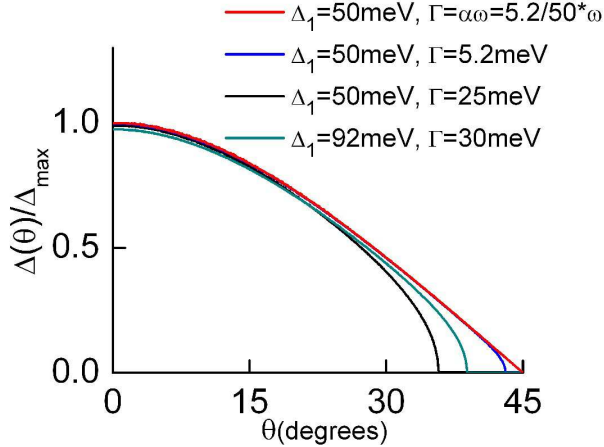}
\caption{Comparison of the angular dependence of the normalized effective gap $\lbrack \Delta (\theta) / \Delta _{max} \rbrack$ on different assumptions for the linewidth $\Gamma$, where $\theta$ denotes the angle relative to the anti-node direction. The assumption of $\Gamma \propto \omega$ at $T \ll T_c$ in Ref.~\cite{Alldredge08} leads to absence of Fermi arc, whereas any finite and constant $\Gamma$ value above $T_c$, as assumed in Ref.~\cite{Norman07}, would lead to a finite Fermi arc length, which corresponds to a finite $\theta$ range over which $\Delta (\theta) = 0$. These two models are not easily reconcilable unless both the temperature and energy dependence of $\Gamma$ switches from one behavior to the other precisely at $T_c$.}
\end{figure}

The aforementioned difficulties encountered in the one-gap scenario strongly suggest that the ground state of cuprate superconductors is unlikely comprised of simply the $d_{x^2-y^2}$-wave superconducting phase. At this point, it seems natural to develop a more quantitative ``two-gap'' model to compare with ARPES and quasiparticle density of states data.

\subsection{Basic concepts and open issues of the two-gap scenario}

The two-gap model is based on substantial empirical evidences and numerical simulations that the ground state of the cuprates may consist of more than one phase for certain range of materials parameters. This occurrence of competing orders (COs) as the result of strong electronic correlations is in fact not unique to the cuprates but has also been found in other correlated electronic systems such as the colossal magnetoresistive (CMR) manganites,~\cite{Dagotto01} fractional quantum Hall liquids~\cite{Finkelstein00,Radzihovsky02} and heavy-fermion superconductors.~\cite{Bauer01,Zapf02} To determine the specific type of COs coexisting with SC for a set of parameters from first principle is a complicated task and is generally carried out numerically.~\cite{Vojta00,Schollwock03} On the other hand, we may investigate the effect of coexisting COs and SC on the low-energy excitations by means of a phenomenological approach that assumes a specific type of CO coexisting with cuprate SC, and then compute the resulting low-energy excitations.~\cite{Yeh07,ChenCT07,Beyer08,YuBL09} Whether the CO is relevant to the specific cuprate can be examined by directly comparing the calculated low-energy excitations with experimental data of ARPES and STS.

To predict the quasiparticle density of states for comparison with tunneling experiments and the quasiparticle spectral density function for comparison with ARPES results, Green function techniques may be employed. The Green function may be calculated by specifying the Hamiltonian of the system to be described. Earlier theoretical approaches to describing the quasiparticle excitation spectra of the cuprates either took the BCS-like Hamiltonian as the unperturbed mean-field state and assumed a disorder-pinned competing order as the perturbation that gives rise to a weak scattering potential for the Bogoliubov quasiparticles,~\cite{Polkovnikov02,ChenCT03,Podolsky03,Capriotti03} or began with the BCS-like Hamiltonian and includes superconducting phase fluctuations in the proper self-energy correction.~\cite{Kwon99,Kwon01} Recent studies improved upon the earlier approximations by incorporating both COs and SC in the mean-field Hamiltonian ${\cal H}_{MF}$ and further including the quantum phase fluctuations associated with the coexisting CO and SC phases in the proper self-energy.~\cite{ChenCT07,Beyer08} Specifically, the approach first evaluated the bare Green function $G_0 (\textbf{k},\omega)$ associated with a mean-field Hamiltonian ${\cal H}_{MF} = {\cal H}_{\rm SC} + {\cal H}_{\rm CO}$ that explicitly included both CO and SC with realistic bandstructure parameters for a given cuprate, where ${\cal H}_{\rm SC}$ is the superconducting Hamiltonian given in Eq.~{\ref{eq:HSC}} for a given pairing potential $\Delta _{\rm SC} (\textbf{k})$, and the explicit form of a given mean-field CO Hamiltonian ${\cal H}_{\rm CO}$ with an energy scale $V_{\rm CO}$ and a wave-vector $\textbf{Q} _1$ for CDW, $\textbf{Q} _2$ for disorder-pinned SDW, $\textbf{Q} _3 = (\pi, \pi)$ for both SDW and DDW may be expressed as follows:~\cite{Beyer09,Teague09,ChenCT07,Beyer08}
\begin{eqnarray}
{\cal H}_{\rm CDW} &= \sum _{{\rm k},\sigma} V_{\rm CDW} \left(
c^{\dagger} _{{\rm k},\sigma} c_{{\rm k}+{\rm Q}_1,\sigma}
+ c^{\dagger} _{{\rm k}+{\rm Q}_1,\sigma} c_{{\rm k},\sigma} \right) \qquad
\qquad \nonumber \\
{\cal H} ^{\rm pinned} _{\rm SDW} &= g^2 \sum _{{\rm k},\sigma} V_{\rm SDW} \left(
c^{\dagger} _{{\rm k},\sigma} c_{{\rm k}+{\rm Q} _2,\sigma}
+ c^{\dagger} _{{\rm k}+{\rm Q} _2,\sigma} c_{{\rm k},\sigma} \right)
\qquad \quad \nonumber \\
{\cal H} _{\rm SDW} &= \sum _{{\rm k},\alpha,\beta} V_{\rm SDW} (\textbf{k}) \left[ c^{\dagger} _{{\rm k}+{\rm Q} _3,\alpha} \sigma _{\alpha \beta} ^3 c_{{\rm k},\beta} \right],
\qquad \qquad \nonumber \\
{\cal H}_{\rm DDW} &= \sum _{{\rm k},\sigma} \frac{1}{2} V_{\rm DDW}
\left( \cos k_x - \cos k_y \right) \left( i c^{\dagger} _{{\rm k}+{\rm Q} _3,\sigma}
c_{{\rm k},\sigma} + h.c. \right) ,
\label{eq:HCO}
\end{eqnarray}
where the coefficient $g$ in ${\cal H}_{\rm SDW} ^{\rm pinned}$ represents the coupling strength between disorder and SDW,~\cite{Polkovnikov02} $\textbf{Q} _1$ is along the CuO$_2$ bonding direction $(\pi , 0)$ or $(0, \pi)$, $\textbf{Q} _2 = \textbf{Q} _1 /2$, $\textbf{Q} _3$ is along $(\pi ,\pi)$ for both SDW~\cite{Schrieffer89} and DDW,~\cite{Chakravarty01} and $\sigma _{\alpha \beta} ^3$ denotes the matrix element $\alpha \beta$ of the $(2 \times 2)$ Pauli matrix $\sigma ^3$. In Eq.~(\ref{eq:HCO}) it is assumed that the COs are density waves and are static because dynamic density waves can be pinned by disorder, and in the latter case the momentum {\bf k} remains a good quantum number as long as the mean free path is much longer than the superconducting coherence length, which is a condition generally satisfied in the cuprates at low temperatures. By diagonalizing ${\cal H}_{\rm MF}$, we obtain the bare Green's function $G_0 (\textbf{k}, \omega)$ for momentum $\textbf{k}$ and energy $\omega$.

To incorporate quantum phase fluctuations, a proper self-energy $\Sigma ^{\ast}$ approximated by the one-loop velocity-velocity correlation may be introduced. In the zero-temperature zero-field limit, the longitudinal phase fluctuations dominate so that $\Sigma ^{\ast} \approx \Sigma ^{\ast} _{\ell}$:~\cite{Kwon99,Kwon01}
\begin{equation}
\Sigma ^{\ast}  _{\ell} \left( \textbf{k}, \omega \right) = \sum _{\textbf{q}}
\left[ m \textbf{v} _g (\textbf{k}) \cdot \hat{\textbf{q}} \right] ^2
C_{\ell} (\textbf{q}) G \left( \textbf{k} - \textbf{q}, \omega \right),
\label{eq:self}
\end{equation}
where the group velocity $\textbf{v}_g$ is given by $\textbf{v}_g = (\partial \xi _{\rm k} / \partial {\rm k})/\hbar$
for $|\textbf{k}| \sim k_F$, and $C_{\ell} (\textbf{q})$ is a coefficient that measures the degree of quantum fluctuations as detailed in Refs.~\cite{ChenCT07,Kwon99,Kwon01}, $\textbf{q}$ is the momentum of quantum phase fluctuations, and $\xi _{\rm k}$ is given by realistic bandstructures.~\cite{ChenCT07,Beyer08} Given $G_0 (\textbf{k}, \omega)$ and $\Sigma ^{\ast} _{\ell} (\textbf{k}, \omega)$ in $(4 \times 4)$ matrices with the basis $(c^{\dagger} _{{\rm k} \uparrow} \ c _{-{\rm k} \downarrow} \ c^{\dagger} _{{\rm k}+{\rm Q} \uparrow} \ c _{-({\rm k}+{\rm Q}) \downarrow})$, the full Green function $G (\textbf{k},\tilde \omega)$ may be obtained through the Dyson's equation:~\cite{ChenCT07,Beyer08}
\begin{equation}
G^{-1} (\textbf{k}, \tilde \omega) = G_0 ^{-1} (\textbf{k}, \omega)
- \Sigma ^{\ast}  _{\ell} ({\rm k}, \tilde \omega),
\label{eq:Dyson}
\end{equation}
where $\tilde \omega = \tilde \omega (\textbf{k}, \omega)$ denotes the energy renormalized by the phase fluctuations.~\cite{ChenCT07,Kwon99} The Dyson's equation in Eq.~(\ref{eq:Dyson}) can be solved self-consistently~\cite{ChenCT07,Beyer08} by first using the mean-field values of $\xi _{\rm k}$ and $\Delta _{\rm SC}$ and choosing an energy $\omega$, then going over the momentum $\textbf{k}$-values in the Brillouin zone by summing over a finite phase space in $\textbf{q}$ near each $\textbf{k}$, and finishing by finding the corresponding fluctuation renormalized quantities $\tilde \xi _{\rm k}$, $\tilde \omega$ and $\tilde \Delta _{\rm SC}$ until the solution to the full Green function $G(\textbf{k}, \tilde \omega)$ converges with an iteration method.~\cite{ChenCT07}
The converged Green function yields the spectral density function $A( \textbf{k}, \omega ) \equiv - \rm{Im} \left[ G(\textbf{k}, \tilde \omega (\textbf{k},\omega)) \right] / \pi$ for comparison with ARPES data~\cite{YuBL09} and the DOS ${\cal N} (\omega) \equiv \sum _{\rm k} A( \textbf{k}, \omega )$ for comparison with STS data.~\cite{ChenCT07,Beyer08} In the case of tunneling experiments, we note that the energy associated with the spectral peaks is given by the maximum value of $\Delta _{\rm eff} (\textbf{k})$, ${\rm max} \lbrace \Delta _{\rm eff} (\textbf{k}) \rbrace \equiv \Delta _{\rm eff} (\textbf{k} _{\rm max}) = \lbrack \Delta _{\rm SC} ^2 (\textbf{k} _{\rm max}) + V_{\rm CO} ^2 (\textbf{k} _{\rm max}) \rbrack ^{1/2}$.~\cite{ChenCT07,Beyer08} This relation in fact holds for all individual wave-vectors $\textbf{k}$, and $\textbf{k} _{\rm max}$ that yields maximum $\Delta _{\rm eff} (\textbf{k})$ is determined by a given set of CO and SC order parameters.

To evaluate temperature-dependent spectral evolution, thermal Green functions are employed in place of the zero-field Green functions. Moreover, it is assumed that the SC gap $\Delta _{\rm SC}$ vanishes at $T_c$ whereas the CO gap $V_{\rm CO}$ vanishes at $T^{\ast}$. In the mean-field limit for the two-gap model, self-consistent gap equations must be solved numerically to account for the temperature-dependent evolution of $V_{\rm CO}$ and $\Delta _{\rm SC}$.~\cite{YuBL09} These effects together with realistic bandstructures of the cuprates have been considered recently in the context of the two-gap model, and favorable comparison with experimental data for both hole- and electron-type cuprates have been confirmed.~\cite{Beyer09,Teague09,ChenCT07,Beyer08,YuBL09} Specifically, the COs in hole-type cuprates appear to be dominated by CDW, disorder-pinned SDW and PDW.~\cite{Beyer09,Teague09,ChenCT07,Beyer08,YuBL09} In contrast, the CO in electron-type cuprates is the commensurate spin density wave (SDW) with a wave-vector $\textbf{Q} = (\pi , \pi)$,~\cite{YuBL09} which is consistent with the findings of neutron scattering experiments,~\cite{Motoyama06} and is also supportive of the findings in recent zero-field STS studies of PLCCO~.\cite{Niestemski07} Moreover, commensurate SDW as the relevant CO in electron-type cuprates leads to $\textbf{k} _{\rm max}$ occurring at the antiferromagnetic ``hot spots'' in the first Brillouin zone, again consistent with experimental findings.~\cite{Matsui05}

\section{A Unified Phenomenology for Unconventional \& Non-Universal Low-Energy Excitations}

The experimental and theoretical status of the cuprates described in previous section suggests that the competing order (or two-gap) scenario is a promising approach to developing a unified phenomenology for cuprate superconductivity. In this section, systematic comparisons of the competing order scenario with the STS, ARPES and magnetization data of both electron- and hole-type cuprates are made as functions of temperature and magnetic field, with favorable agreements suggesting a feasible phenomenology that unifies most empirical findings to date.

\subsection{Temperature-dependent tunneling spectra in zero fields}

To begin, we consider the $T$-evolution of the quasiparticle tunneling spectra in zero fields by means of the Green function technique and the mean-field Hamiltonian ${\cal H} _{MF} = {\cal H}_{\rm SC} + {\cal H} _{\rm CO}$, where ${\cal H}_{\rm SC}$ and ${\cal H}_{\rm CO}$ are given in Eqs.~(\ref{eq:HSC}) and (\ref{eq:HCO}), respectively. For coexisting $d_{x^2-y^2}$-wave SC and a specific CO, the zero-field quasiparticle spectra ${\cal N}(\omega)$ and $A(\textbf{k},\omega)$ at $T$ = 0 can be fully determined by the parameters $\Delta _{\rm SC}$, $V_{\rm CO}$, $\textbf{Q}$, $\delta \textbf{Q}$ (the linewidth of the wave-vector), $\Gamma _{\rm k}$ (the quasiparticle linewidth), and $\eta$ (the magnitude of quantum phase fluctuations).~\cite{ChenCT07,Beyer08} In Figs.~8(a) and 8(b) exemplified STS data for optimally doped cuprate superconductors Y-123 and La-112 together with the corresponding theoretical fittings are shown for comparison. Using the aforementioned theoretical analysis, a consistent account for the $T$-dependent quasiparticle tunneling spectra in both hole- and electron-type cuprates can be obtained by assuming Fermi-surface nested CDW~\cite{Wise08,Beyer09,Beyer08} as the CO in the hole-type cuprates and commensurate SDW as the CO in the electron-type cuprates.~\cite{Teague09,YuBL09} Specifically, for hole-type cuprates such as in the spectra of Y-123, the sharp peaks and satellite ``hump'' features at $T \ll T_c$ in Fig. 8(a) are associated with $\omega = \pm \Delta _{\rm SC}$ and $\omega = \pm \Delta _{\rm eff}$, respectively, where $\Delta _{\rm eff} \equiv \lbrack (\Delta _{\rm SC})^2 + (V_{\rm CO})^2]^{1/2}$ is an effective excitation gap. Hence, it may be concluded that the condition $V_{\rm CO} > \Delta _{\rm SC}$ in hole-type cuprates is responsible for the appearance of the satellite features at $T \ll T_c$ and the PG phenomena at $T^{\ast} > T > T_c$~\cite{Beyer09,Beyer08,YuBL09}. In contrast, the condition $V_{\rm CO} < \Delta _{\rm SC}$ in electron-type cuprates, as exemplified in Fig. 8(b), is responsible for only one set of characteristic features at $\omega = \pm \Delta _{\rm eff}$ and the absence of PG above $T_c$. In contrast to the strong spatially inhomogeneous spectra in Bi-2212, it is also worth noting that both optimally doped Y-123 and La-112 exhibit relatively long-range spectral homogeneity,~\cite{Beyer09,Teague09} as manifested by the energy histograms shown in Figs.~8(c) and 8(d). Hence, the exemplified theoretical and experimental comparions in Figs.~8(a) and 8(b) are in fact representative of the general behavior found in these cuprates.

\begin{figure}[!]
\centering
\includegraphics[keepaspectratio=1,width=8cm]{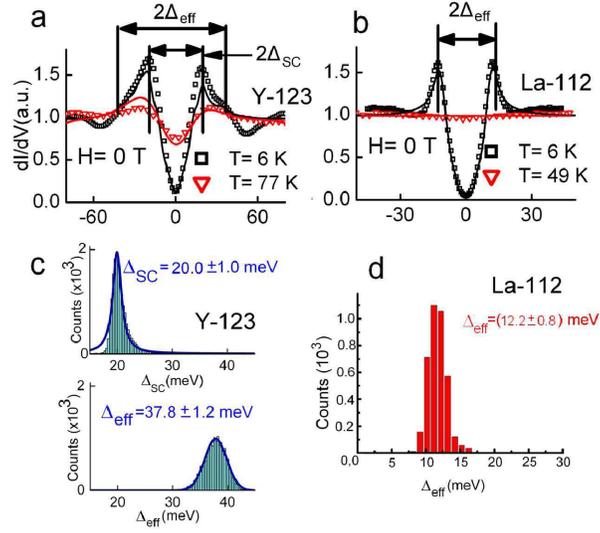}
\caption{Implication of CO from zero-field STS in Y-123 and La-112: (a) Normalized zero-field tunneling spectra of Y-123 taken at $T$ = 6 K (black) and 77 K (red). The solid lines represent fittings to the $T$ = 6 and 77 K spectra by assuming coexisting SC and CDW, with fitting parameters of $\Delta _{\rm SC}$ = 20 meV, $V_{\rm CDW}$ = 32 meV and $\textbf{Q}_{\rm CDW} = (0.25 \pi \pm 0.05 \pi, 0) / (0, 0.25 \pi \pm 0.05 \pi)$, following Refs.~\cite{ChenCT07,Beyer08}. (b) Normalized zero-field tunneling spectra of La-112 taken at T = 6 K (black) and 49 K (red). The solid lines represent fittings to the $T$ = 6 and 49 K spectra by assuming coexisting SC and SDW, with fitting parameters $\Delta _{\rm SC}$ = 12 meV, $V_{\rm SDW}$  = 8 meV, and $\textbf{Q}_{\rm SDW} = (\pi , \pi)$, following Refs.~\cite{ChenCT07,Beyer08}. (c) Energy histogram of $\Delta _{\rm SC}$ (upper panel) and $\Delta _{\rm eff}$ (lower panel) in Y-123 over a $(95 \times 95) {\rm nm}^2$ area. (d) Energy histogram of $\Delta _{\rm eff}$ in La-112 over a $(64 \times 64) {\rm nm}^2$ area.}
\end{figure}

By extending the above CO analysis to the STS data of hole-type cuprates with different doping levels  ($\delta$),~\cite{Beyer08} the resulting $\Delta _{\rm SC} (\delta)$ generally follows the same non-monotonic dependence of $T_c (\delta)$, whereas $V_{\rm CO} (\delta )$ increases with decreasing $\delta$, consistent with the general trend of the zero-field PG temperature in hole-type cuprates.~\cite{Beyer08} These findings are summarized in Fig.~9(a). Further analysis of the ratios of $\Delta _{\rm SC} (\delta)/(k_B T_c)$, $V _{\rm CO} (\delta)/(k_B T^{\ast})$ and $\Delta _{\rm eff} (\delta)/(k_B T_c)$ yields the results shown in Fig.~9(b), where the PG temperatures $T^{\ast}$ are obtained from Refs.~\cite{LeeWS07,Kanigel06}. Interestingly, in the two-gap scenario the ratio $\Delta _{\rm SC} (\delta)/(k_B T_c)$ increase slightly from $\sim 3.5$ in the overdoped limit to $\sim 4.5$ in the underdoped limit. The former is close to the BCS value for $s$-wave superconductivity and the latter for $d_{x^2-y^2}$-wave superconductivity. Noting that the pairing symmetry of Y-123 evolves from nearly pure $d_{x^2-y^2}$-wave to $d_{x^2-y^2}+s$-wave with increasing doping,~\cite{Yeh01,Masui03} these values are in fact reasonable. In contrast, the ratio $V _{\rm CO} (\delta)/(k_B T^{\ast})$ is found to be nearly a constant $\sim 2.5$ for all doping levels. On the other hand, the ratio $\Delta _{\rm eff} (\delta)/(k_B T_c)$ commonly considered by the one-gap model shows a rapid increase with decreasing doping. These findings are consistent with other reports, confirming the notion that the effective gap $\Delta _{\rm eff}$ considered by the one-gap model is not the superconducting gap and is in fact a convoluted excitation gap of admixtures of Bogoliubov quasiparticles and collective modes.

\begin{figure}[!]
\centering
\includegraphics[keepaspectratio=1,width=12cm]{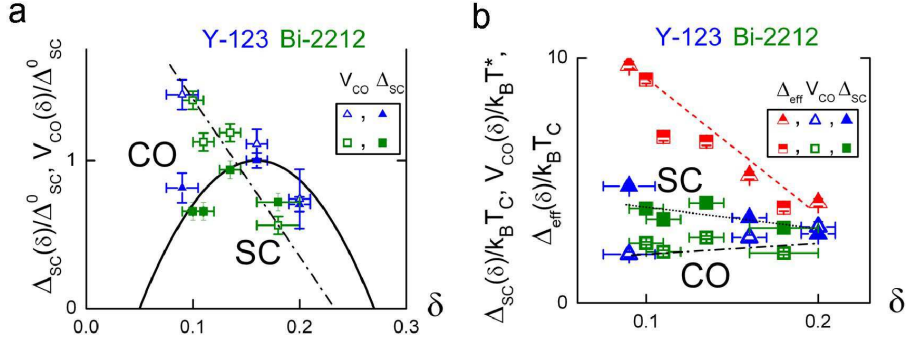}
\caption{Doping dependent SC and CO gaps in Y-123 and Bi-2212: (a) Comparison of the doping dependence of $\Delta _{\rm SC}$, $V_{\rm CO}$ and $T_c$ for hole-type cuprates Y-123 and Bi-2212, where $\Delta _{\rm SC} (\delta)$ and $V_{\rm CO} (\delta)$ are normalized to their respective SC gaps at the optimal doping, $\Delta ^0 _{\rm SC}$, and $T_c(\delta)$ is normalized to the optimal doping value $T_c ^0$.~\cite{Beyer08} (b) Comparison of the doping dependent ratios of $\Delta _{\rm SC} /(k_B T_c)$, $V _{\rm CO} /(k_B T^{\ast})$ and $\Delta _{\rm eff} /(k_B T_c)$.}
\end{figure}

In addition to the manifestation of two energy scales in the tunneling spectra, the presence of LDOS modulations with energy-independent wave-vectors also provides strong evidence for the existence of competing orders. As exemplified in Figs.~6(a)-(c) for zero-field Y-123 tunneling conductance maps taken at constant energies of $\omega = -23$ meV, $-33$ meV and $-53$ meV, apparent density-wave modulations are visible. The presence of density waves can be further confirmed by means of Fourier-transformed (FT) LDOS studies.~\cite{Beyer09} Similar LDOS modulations have also been found in the Bi-2201 system as functions of the hole doping level~\cite{Boyer07,Wise08} and have been attributed to Fermi-surface nested CDW with a decreasing CDW wave-vector upon decreasing doping. The finding of Fermi surface-nested collective modes is consistent with the predictions of the two-gap model as the more favorable low-energy excitations.~\cite{Beyer08} As detailed in Ref.~\cite{Beyer09} and exemplified in Figs.~10(a) and 10(b) for the FT-LDOS data at $\omega = -8$ meV and $-13$ meV, respectively, systematic analysis of the energy dependence of the FT-LDOS, $F(\textbf{k},\omega)$, reveals two types of diffraction spots in the momentum space.~\cite{Beyer09} One type of spots are strongly energy dependent as shown in Fig.~5(c), which may be attributed to elastic quasiparticle scattering interferences as described by the octet model.~\cite{McElroy05,ChenCT03} The other type of spots are nearly energy-independent, which are circled in Figs.~10(a) and 10(b) for FT-LDOS at $H$ = 0, with the corresponding momentum $vs.$ energy dependence shown in Fig.~5(d).~\cite{Yeh09,Beyer09} More specifically, in addition to the reciprocal lattice vectors and the $(\pi, \pi)$ resonance mode,~\cite{Beyer09} there are two sets of nearly energy-independent wave-vectors along $(\pi ,0)/(0,\pi )$, which are tentatively denoted as $\textbf{Q}_{\rm PDW}$ and $\textbf{Q}_{\rm CDW}$; and one set of energy-independent wave-vector along $(\pi , \pi )$, which is denoted by $\textbf{Q}_{SDW}$. Quantitatively, $\textbf{Q}_{\rm PDW} = \lbrack (0.56 \pi \pm 0.06 \pi) /a_1,0 \rbrack$ and $\lbrack 0, ((0.56 \pi \pm 0.06 \pi)/a_2 \rbrack$, $\textbf{Q}_{\rm CDW} = \lbrack (0.28 \pi \pm 0.02 \pi) /a_1, 0 \rbrack$ and $\lbrack 0, (0.28 \pi \pm 0.02 \pi) /a_2 \rbrack$, and $\textbf{Q}_{\rm SDW} = \lbrack (0.15 \pi \pm 0.01 \pi)/a_1, (0.15 \pi \pm 0.01 \pi)/a_2 \rbrack$.~\cite{Beyer09} Here $a_1 = 0.383$ nm and $a_2 = 0.388$ nm are the lattice constants of optimally doped Y-123.

\begin{figure}[!]
\centering
\includegraphics[keepaspectratio=1,width=12cm]{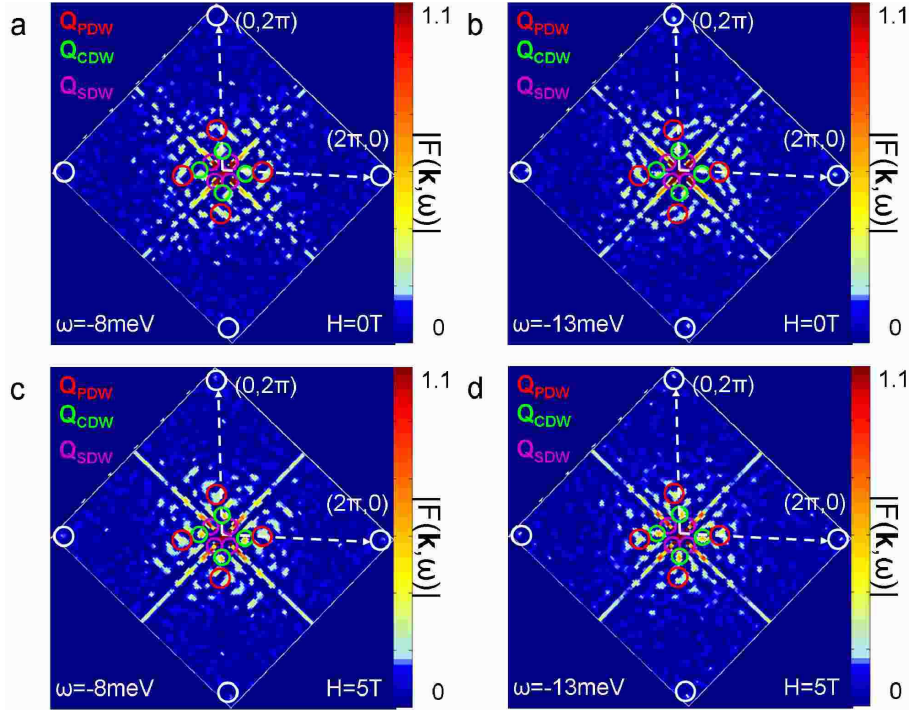}
\caption{The intensity of the FT-LDOS of Y-123, $|F(\textbf{k}, \omega )|$, is shown in two-dimensional momentum space for (a) $H$ = 0 and $\omega = -8$ meV, (b) $H$ = 0 and $\omega = -13$ meV, (c) $H$ = 5 T and $\omega = -8$ meV, and (d) $H$ = 5 T and $\omega = -13$ meV.~\cite{Beyer09} Comparing (a) with (b) and (c) with (d), three sets of $\omega$-independent wave-vectors in addition to the reciprocal lattice constants and the $(\pi , \pi)$ resonance may be identified: $\textbf{Q}_{\rm PDW}$ and $\textbf{Q}_{\rm CDW}$ along $(\pi ,0)/(0, \pi)$ and $\textbf{Q}_{\rm SDW}$ along $(\pi,\pi)$, which are circled for clarity.}
\end{figure}

\subsection{Spatially resolved vortex-state quasiparticle tunneling spectra}

An alternative way of verifying the feasibility of the CO scenario is to introduce vortices because the suppression of SC inside vortices may unravel the spectroscopic characteristics of the remaining CO. As exemplified in Fig.~11(a) for a set of intra- and inter-vortex spectra taken on Y-123 at $H$ = 2 T and $T$ = 6 K, and in Fig. 11(b) for a set of intra- and inter-vortex spectra taken on La-112 at $H$ = 1 T, the quasiparticle spectra near the center of each vortex exhibit pseudogap (PG)-like features, which is in stark contrast to theoretical predictions for a sharp zero-energy peak around the center of the vortex core had SC been the sole order in the ground state.~\cite{Caroli64,Hess90,Franz98} Interestingly, for both Y-123 and La-112, the respective PG energy inside vortices is comparable to the CO energy $V_{\rm CO}$ derived from the zero-field fittings in Figs.~8(a) and 8(b). Additionally, a subgap feature at $\Delta ^{\prime} < \Delta _{\rm SC}$ is found inside the vortex cores of Y-123. To investigate how quasiparticle spectra evolve with magnetic field, spatially resolved spectroscopic studies at varying fields were made for both Y-123 and La-112 at T = 6 K.~\cite{Beyer09,Teague09} In Figs.~11(c) and 11(d) the field dependent energy histograms of Y-123 and La-112 are shown. In the case of Y-123, a strong spectral shift from SC at $\omega = \Delta _{\rm SC}$ to PG at $\omega = V_{\rm CO} > \Delta _{\rm SC}$ is seen with increasing $H$, together with the appearance of a third subgap (SG) feature at $\omega = \Delta ^{\prime} < \Delta _{\rm SC}$. In contrast, for La-112 the energy histogram at each magnetic field can be fit by a Lorentzian functional form with a peak energy at $\Delta _{\rm eff}(H)$, which decreases slightly with increasing $H$.~\cite{Teague09} Additionally, there is an apparent low energy ``cutoff'' at $V_{\rm CO} (\sim 8 {\rm meV}) < \Delta _{\rm SC} (\sim 12 {\rm meV})$ for all energy histograms. Neither the histograms of Y-123 nor those of La-112 exhibit any spectral peak at $\omega = 0$, which are in sharp contrast to the steady increase of the spectral weight at $\omega = 0$ for conventional type-II superconductors.~\cite{Caroli64,Hess90} These field-revealed energy gaps $\Delta ^{\prime}$ and $V_{\rm CO}$ and the $\omega$-independent wave-vectors for the LDOS modulations in the cuprates cannot be explained by assuming pure SC in the ground state. Moreover, the observation of apparent collective modes $\textbf{Q} _{\rm CDW}$ and $\textbf{Q} _{\rm PDW}$ in Y-123~\cite{Beyer09}, $\textbf{Q} _{\rm PDW}$ in Bi-2212~\cite{Hoffman02,Howald03} and $\textbf{Q} _{\rm CDW}$ in Bi-2201~\cite{Boyer07,Wise08} may also explain the dichotomy of quasiparticle coherence (i.e. incoherent anti-nodal quasiparticles and coherent nodal quasiparticles)~\cite{ChenCT07,ZhouXJ04} and the breakdown of the octet model for quasiparticle scattering momentum near the anti-nodes.~\cite{Hanaguri09}

\begin{figure}[!]
\centering
\includegraphics[keepaspectratio=1,width=12cm]{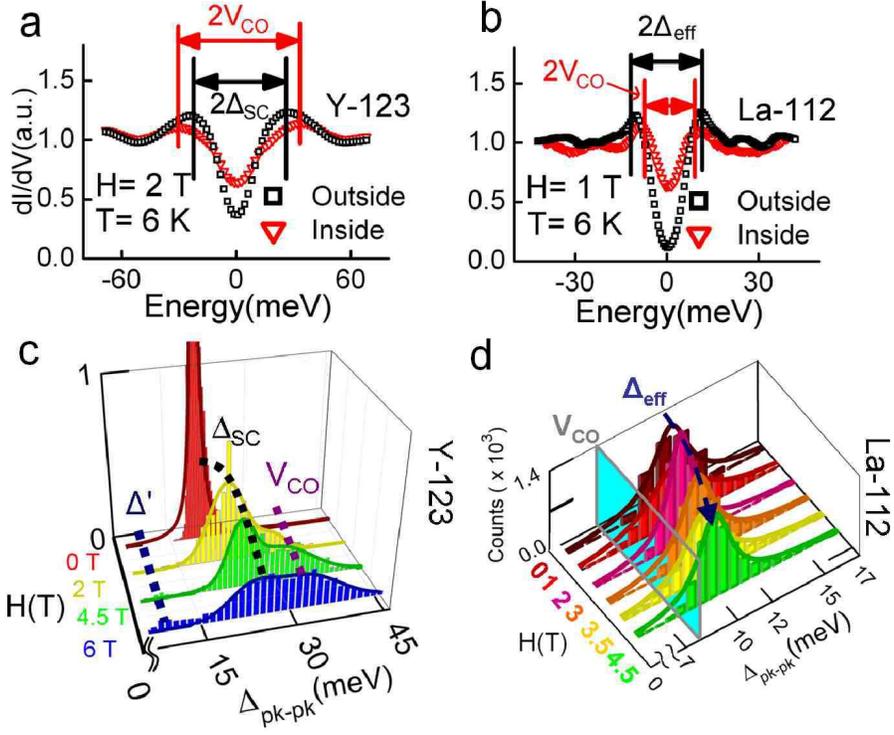}
\caption{(a) Spatially averaged intra- and inter-vortex spectra of Y-123 for $T$ = 6 K and $H$ = 2 T,~\cite{Beyer09} showing PG features inside the vortex, with a PG energy larger than $\Delta _{\rm SC}$ and consistent with the $V_{\rm CDW}$ value derived from fitting the zero-field spectra in Fig.~8(a). (b) Spatially averaged intra- and inter-vortex spectra of La-112 for $T$ = 6 K and $H$ = 1 T,~\cite{Teague09} showing PG features inside the vortex, with a PG energy smaller than $\Delta _{\rm eff}$ and consistent with the $V_{\rm SDW}$ value derived from fitting the zero-field spectra in Fig.~8(b). (c) Magnetic field-dependent spectral evolution in cuprate superconductors at $T$ = 6 K, showing energy histograms derived from the quasiparticle tunneling spectra of Y-123 for $H$ = 0, 2, 4.5, and 6 T. A steady spectral shift from $\Delta _{\rm SC}$ to $V_{\rm CO}$ and $\Delta ^{\prime}$ is found with increasing $H$.~\cite{Beyer09} (d) Magnetic field dependence of the characteristic energies $\Delta_{\rm eff}$, $\Delta _{\rm SC}$ and $V_{\rm CO}$ in La-112 for $H$ = 0, 1, 2, 3, 3.5 and 4.5 T. Each histogram can be fit with a Lorentzian functional form. The peak position of the Lorentian is identified as $\Delta_{\rm eff} (H)$, and the low-energy cutoff of the histogram is identified as $V_{\rm CO}$. Empirically, $V_{\rm CO}$ is nearly constant whereas $\Delta_{\rm eff} (H)$ decreases slightly with increasing $H$.~\cite{Teague09}}
\end{figure}

The FT-LDOS analysis of the vortex-state spectra yield results similar to the zero-field data. As manifested in Figs.~10(c)-(d) for FT-LDOS taken at $H$ = 5 T and $\omega = -8$ meV and $-13$ meV, respectively, similar energy-independent wave-vectors $\textbf{Q}_{\rm PDW}$ and $\textbf{Q}_{\rm CDW}$ along $(\pi,0)/(0,\pi)$ and $\textbf{Q}_{\rm SDW}$ along $(\pi, \pi)$ are found, in addition to the strongly dispersive modes. However, further inspection of the intensity of the Fourier-transformed conductance $|F(\textbf{k}, \omega )|$~\cite{Beyer09} reveal interesting evolution with $\omega$ and $H$. As shown in Figs.~12(a)-(d), the intensities of $|F(\textbf{k}, \omega )|$ associated with the collective modes (at $|\textbf{k}|$ = $|\textbf{Q} _{\rm CDW}|$ and $|\textbf{Q} _{\rm PDW}|$ for $\textbf{k} \parallel (\pi ,0)$ and at $|\textbf{k}|$ = $|\textbf{Q} _{\rm SDW}|$ for $\textbf{k} \parallel (\pi ,\pi)$) on the $|\textbf{k}|$-vs.-$\omega$ plane are all enhanced by magnetic fields. However, the non-motonic intensity evolution of $|F(\textbf{k}, \omega )|$ with $\omega$ for all three collective modes is not understood. For comparison, in Fig.~12(e) the $|\textbf{q}|$-vs.-$\omega$ relations of $\textbf{Q} _{\rm PDW}$, $\textbf{Q} _{\rm CDW}$ and $\textbf{Q} _{\rm SDW}$ are reproduced, together with a dispersive wave-vector associated with quasiparticle scattering interferences~\cite{McElroy05,ChenCT03} along $(\pi ,\pi)$, which is denoted as $\textbf{q}_7$ in Ref.~\cite{McElroy05} and Fig.~5(a). Overall, the dispersion relations for the quasiparticle scattering intereference modes shown in Fig.~12(e) and Figs.~5(c) for Y-123 are in good agreement with the experimental results found in Bi-2212~\cite{McElroy05}. Additionally, the $\textbf{Q} _{\rm PDW}$ found in Y-123 is consistent with the wave-vector of the checkerboard patterns observed in the vortex cores of Bi-2212.~\cite{Hoffman02} On the other hand, the modes $\textbf{Q} _{\rm CDW}$ and $\textbf{Q} _{\rm SDW}$ have not been reported in the Bi-2212 systems.

\begin{figure}[!]
\centering
\includegraphics[keepaspectratio=1,width=12cm]{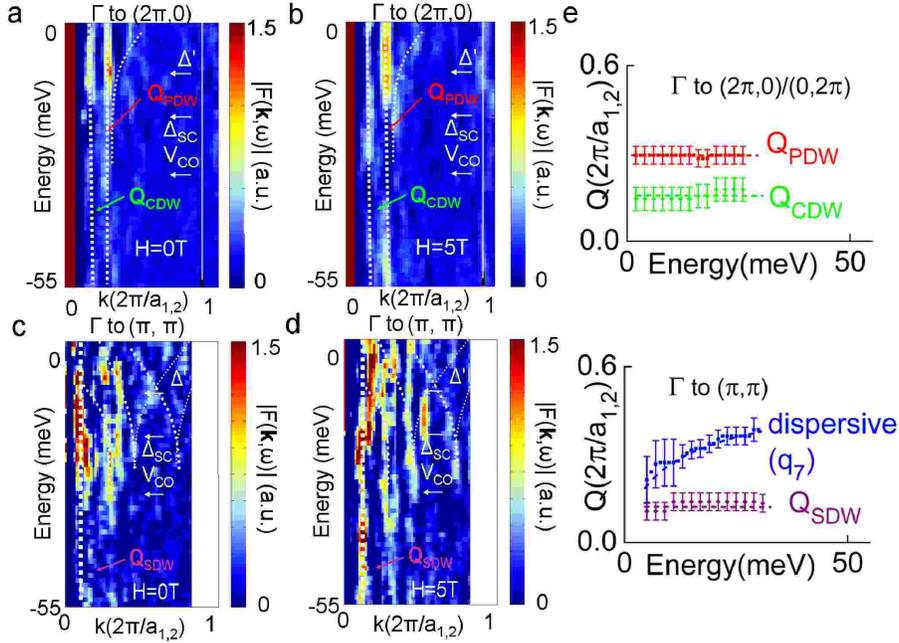}
\caption{(a) The $\omega$-dependence of $|F(\textbf{k}, \omega )|$ at $H$ = 0 is plotted against $\textbf{k} \parallel (\pi ,0)$, showing $\omega$-independent modes (bright vertical lines) at $\textbf{k} = \textbf{Q}_{\rm PDW}$ and $\textbf{Q}_{\rm CDW}$. (b) The $\omega$-dependence of $|F(\textbf{k}, \omega )|$ at $H$ = 5 T is plotted against $\textbf{k} \parallel (\pi ,0)$, showing field-enhanced intensity at $\textbf{k} = \textbf{Q}_{\rm PDW}$ and $\textbf{Q}_{\rm CDW}$. (c) $|F(\textbf{k}, \omega )|$ at $H$ = 0 is plotted against $\textbf{k} \parallel (\pi ,\pi)$, showing $\omega$-independent modes (bright vertical lines) at $\textbf{k} = \textbf{Q}_{\rm SDW}$ together with other dispersive modes. (d) $|F(\textbf{k}, \omega )|$ at $H$ = 5 T is plotted against $\textbf{k} \parallel (\pi ,\pi)$, showing field-enhanced intensity at $\textbf{k} = \textbf{Q}_{\rm SDW}$. (e) Upper panel: Momentum $(|\textbf{q}|)$ vs. energy ($\omega$) for $\textbf{q}$ = $\textbf{Q}_{\rm PDW}$ and $\textbf{Q}_{\rm CDW}$ along $(\pi ,0)$. Lower panel: $|\textbf{q}|$ vs. $\omega$ for $\textbf{q}$ = $\textbf{Q}_{\rm SDW}$ along $(\pi,\pi)$. A representative dispersive wave-vector $\textbf{q}_7$ along $(\pi,\pi)$ due to quasiparticle scattering interferences~\cite{McElroy05,ChenCT03} is also shown in the lower panel for comparison.}
\end{figure}

The occurrence of two energy scales $V_{\rm CO} (> \Delta _{\rm SC})$ and $\Delta ^{\prime} (< \Delta _{\rm SC})$ inside vortices and the field-enhanced energy-independent wave-vectors $\textbf{Q}_{\rm PDW}$, $\textbf{Q}_{\rm CDW}$ and $\textbf{Q}_{\rm SDW}$ in Y-123 strongly suggest that the vortex-state quasiparticle tunneling spectra in Y-123 cannot be explained by simple Bogoliubov quasiparticle scattering interferences alone~\cite{McElroy05,ChenCT03,Hanaguri09}. On the other hand, these findings may be compared with the scenario of coexisting CO's and SC.~\cite{Kivelson03,Boyer07,Wise08,Beyer09,Teague09,Demler01,Polkovnikov02,ChenHD02,ChenHD04,LiJX06,ChenCT07,Beyer08} In particular, the energy scale $V_{\rm CO}$ and the wave-vector $\textbf{Q}_{\rm CDW}$ manifested in the vortex-state spectra are consistent with the CO parameters derived from the Green function analysis of the zero-field data.~\cite{Beyer09} For comparison, the vortex-state STS studies on La-112 also revealed PG-like features inside vortices,~\cite{Teague09} except that the PG energy in La-112 is {\it smaller} than $\Delta _{\rm SC}$, as shown in Figs.~11(b) and 11(d). This finding may also be interpreted as a CO being revealed inside the vortex core upon the suppression of SC. Furthermore, the smaller energy associated with the PG-like features inside vortices of La-112 is consistent with the {\it absence} of PG above $T_c$ in electron-type cuprate superconductors.~\cite{Teague09} Thus, the contrasts between the vortex-state quasiparticle spectra of hole-type and those of the electron-type cuprates may be attributed to the different magnitude of $V_{\rm CO}$ relative to $\Delta _{\rm SC}$.~\cite{Beyer09,Teague09}

\subsection{Application of the competing order scenario to ARPES data}

In addition to the effect of COs on the quasiparticle tunneling spectra of hole- and electron-type cuprates, how COs influence ARPES may be investigated using the Green function technique, and the internal consistency of the ARPES analysis with tunneling spectroscopy may be evaluated.~\cite{YuBL09} It is found that the appearance of the low-energy PG in hole-type cuprates may be correlated with the appearance of the Fermi arc above $T_c$ and below the PG temperature $T^{\ast}$~\cite{Damascelli03,LeeWS07} within the CO scenario,~\cite{YuBL09} and that the spectral density functions derived from fitting the temperature and doping dependent ARPES data also yield quasiparticle spectra consistent with spatially averaged STS results at $T \ll T_c$.~\cite{YuBL09} As exemplified in Figs.~13(a)-(c) for three Bi-2212 samples of different doping levels,~\cite{YuBL09} the Fermi arc as a function of the quasiparticle momentum $\textbf{k}$, temperature ($T$) and doping level ($\delta$) in Bi-2212~\cite{LeeWS07} may be explained consistently by assuming incommensurate CDW with $V _{\rm CDW} > \Delta _{\rm SC}$ and $\textbf{Q} _{\rm CDW} \parallel (\pi,0)/(0,\pi)$ as the relevant competing order. Additionally, the Fermi arc length is found to follow a universal dependence on the reduced temperature $(T/T^{\ast})$ for all doping levels.~\cite{YuBL09} That is,
\begin{equation}
{\rm arc \ length} \sim \frac{2}{\pi} \left( \frac{\pi}{2} - 2.5 \delta Q \right) ; \qquad \delta Q \sim 0.21 \pi \left( 1- \frac{T}{T^{\ast}} \right) ^{0.53}.
\label{eq:arc}
\end{equation}
The universal trend given in Eq.~(\ref{eq:arc}) is consistent among all Fermi arc data of hole-type cuprates.~\cite{LeeWS07,Kanigel06} Similarly, the $\textbf{k}$- and $T$-dependence of the effective gap $\Delta _{\rm eff}$ and the absence of Fermi arcs in electron-type cuprates (e.g. $\rm Pr_{0.89}LaCe_{0.11}CuO_4$, denoted as PLCCO)~\cite{Matsui05} can also be explained by incorporating commensurate SDW with $V _{\rm SDW} < \Delta _{\rm SC}$ and $\textbf{Q} _{\rm SDW} = (\pi,\pi)$ into spectral characteristics,~\cite{YuBL09} as shown in Figs.~13(d)-(f). Hence, the CO scenario can provide unified phenomenology for both ARPES and STS data of hole- and electron-type cuprates as functions of temperature and doping level.

\begin{figure}[!]
\centering
\includegraphics[keepaspectratio=1,width=13cm]{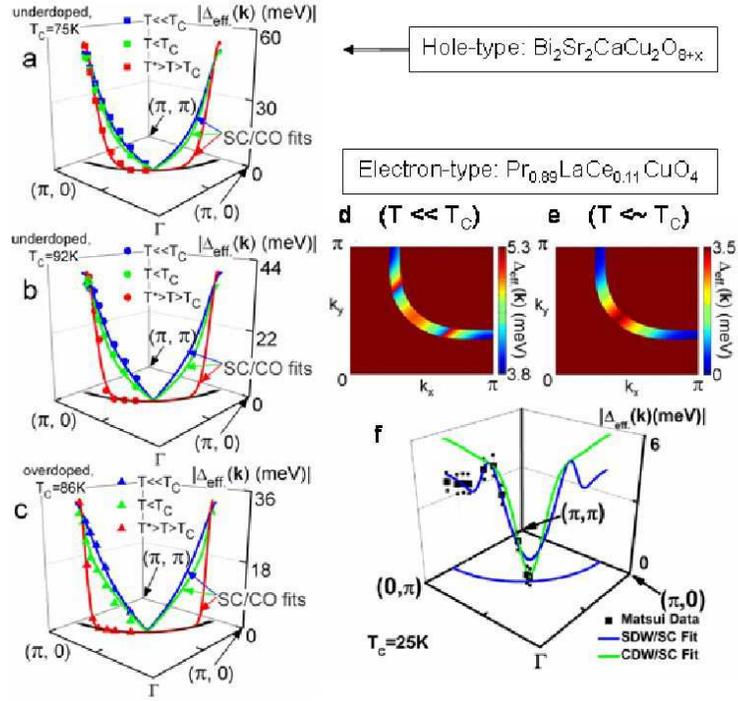}
\caption{((a) -– (c) Theoretical fittings (solid color curves)~\cite{YuBL09} to the momentum ($\textbf{k}$) dependent effective excitation gap $\Delta _{\rm eff}$ (color symbols) determined from ARPES spectra on Bi-2212 for three different doping levels and at three temperatures.~\cite{LeeWS07} The $d_{x^2-y^2}$-SC/CDW scenario~\cite{YuBL09} agrees with experimental findings:~\cite{LeeWS07} Below $T_c$, $\Delta _{\rm eff}(\textbf{k})$ only vanishes at the nodal point, whereas above $T_c$ and below $T^{\ast}$, $\Delta _{\rm eff}(\textbf{k})$ vanishes over an ``arc''-shaped of wave-vectors. In contrast, the $\textbf{k}$- and $T$-dependence of $\Delta _{\rm eff}(\textbf{k})$ in electron-type cuprates differs from that in the hole-type cuprate $\rm Pr_{0.89}LaCe_{0.11}CuO_4$, as exemplified in (d) –- (f).~\cite{Matsui05}}
\end{figure}

\subsection{Competing orders as the physical origin of spatially inhomogeneous tunneling spectra in Bi-2212}

In the context of spatial homogeneity of quasiparticle spectra, STS studies of optimally doped Y-123 have shown  spatially homogeneous $\Delta _{\rm SC}$ and less homogeneous $V_{\rm CO}$, as exemplified in Fig.~8(c).~\cite{Beyer09,Yeh01} Similarly, STS studies of the optimally doped La-112 also reveal highly homogeneous spatial distributions of $\Delta _{\rm eff}$ as shown in Fig.~8(d).~\cite{ChenCT02,Teague09} These results are in stark contrast to the findings of a highly inhomogeneous effective quasiparticle excitation gap $\Delta _{\rm eff}$ in Bi-2212.~\cite{McElroy05,Lang02,LeeJ06} The strong inhomogeneity in $\Delta _{\rm eff}$ of the Bi-2212 system may be attributed to its extreme two-dimensional nature so that disorder has more significant effect (such as pinning effect) on the low-energy quasiparticle excitations. If the inhomogeneous effective excitation gap $\Delta _{\rm eff}$ in Bi-2212 is primarily due to inhomogeneous $V_{\rm CO}$ and if the CO is associated with incommensurate CDW, realistic energy histograms of $\Delta _{\rm eff} (\delta)$~\cite{McElroy05,LeeJ06} may be employed to simulate two-dimensional spatial maps of $\Delta _{\rm eff}$, LDOS and FT-LDOS. It is found that such assumptions yield spectral maps consistent with empirical observation,~\cite{LeeJ06} as exemplified in Fig.~14.~\cite{Yeh09,LeeSP09} On the other hand, if the inhomogeneity in $\Delta _{\rm eff}$ were primarily associated with that in $\Delta _{\rm SC}$, the corresponding LDOS near $\omega = 0$ would have become strongly varying in space,~\cite{LeeSP09} which contradicts the experimental findings of highly homogeneous spectra near $\omega = 0$.~\cite{McElroy05,Lang02} It is further noted that COs must be included in the simulations of the LDOS and FT-LDOS of Bi-2212 to achieve results consistent with experimental observation,~\cite{ChenCT03,LeeSP09} because the assumption of pure SC as in the one-gap model cannot account for various spectral characteristics, including the periodic modulations in the LDOS and the non-dispersive wave-vectors along the $(\pi,0)$ and $(0,\pi)$ directions in the FT-LDOS.

\begin{figure}[!]
\centering
\includegraphics[keepaspectratio=1,width=13cm]{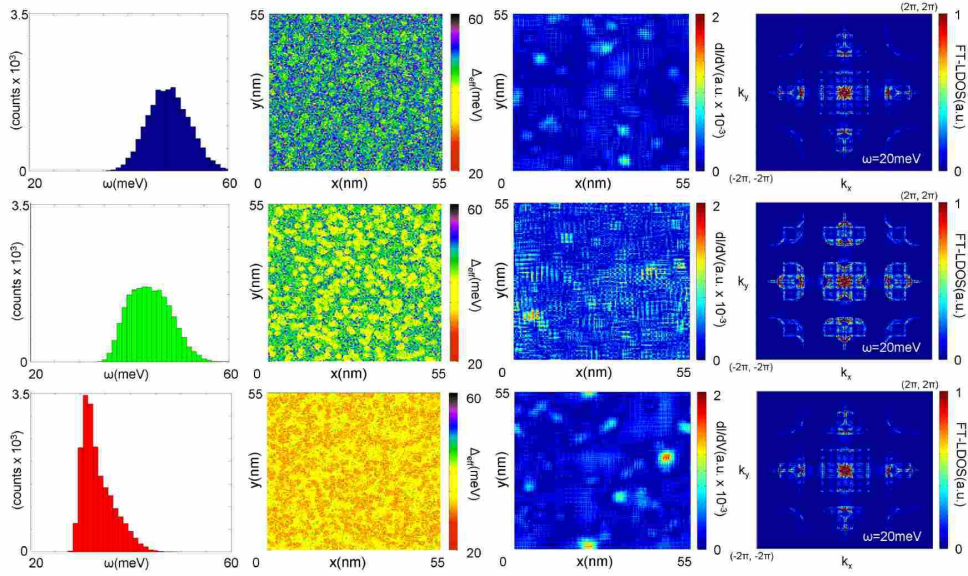}
\caption{Simulations for doping-dependent quasiparticle spectral characteristics over a $(55 \times 55) \rm nm^2$ area based on the CO scenario~\cite{ChenCT07,Beyer08,ChenCT03} and the conjecture of coexisting homogeneous $\Delta _{\rm SC}(\delta)$ and inhomogeneous $V_{\rm CO}(\delta)$ associated with 50 impurities.~\cite{Yeh09,LeeSP09} Here $\delta$ = 0.12 (first row), 0.15 (second row), and 0.19 (third row). First column: histograms of $\Delta _{\rm SC}(\delta)$; second column, $\Delta _{\rm eff}(\delta)$ maps; third column, LDOS maps at $\omega$ = 20 meV; fourth column, FT-LDOS maps at $\omega$ = 20 meV.~\cite{Yeh09}}
\end{figure}

\subsection{Strong field-induced quantum phase fluctuations}

As described earlier, cuprate superconductors are doped Mott insulators with strong electronic correlation that can result in a variety of competing orders in the ground state.~\cite{Kivelson03,Sachdev03,Yeh02a,Zhang97} Therefore, significant quantum fluctuations and reduced SC stiffness are expected in the cuprates because of the existence of multiple channels of low-energy excitations. Macroscopically, all cuprate superconductors are known to be extreme type-II superconductors,~\cite{Blatter94,FisherDS91}, which implies weak SC stiffness associated with these doped Mott insulators. Moreover, external variables such as temperature ($T$) and applied magnetic field ($H$) can vary the interplay of SC and CO, such as inducing or enhancing the CO~\cite{Lake01,ChenY02} at the price of more rapid suppression of SC, thereby leading to weakened superconducting stiffness and strong thermal and field-induced fluctuations.~\cite{Beyer07,Zapf05,Blatter94,FisherDS91} On the other hand, the quasi two-dimensional nature of the cuprates can also lead to quantum criticality in the limit of decoupling of CuO$_2$ planes. Indeed, recent studies have demonstrated experimental evidence from {\it macroscopic} magnetization measurements for field-induced quantum fluctuations among a wide variety of cuprate superconductors with different {\it microscopic} variables such as the doping level ($\delta$) of holes or electrons, and the number of CuO$_2$ layers per unit cell ($n$).~\cite{Chakravarty04} It may be suggested that the manifestation of strong field-induced quantum fluctuations is consistent with a scenario that all cuprates are in close proximity to a quantum critical point (QCP).~\cite{Kotliar96}

To investigate the effect of quantum fluctuations on the vortex dynamics of cuprate superconductors, vortex phase diagrams for different cuprates were studied at $T \to 0$ to minimize the effect of thermal fluctuations, and the magnetic field was applied parallel to the CuO$_2$ planes ($H \parallel ab$) to minimize the effect of random point disorder.~\cite{Beyer07} More specifically, the rationale for having $H \parallel ab$ is that the intrinsic pinning effect of layered CuO$_2$ planes generally dominates over the pinning effects of random point disorder, so that the commonly observed glassy vortex phases associated with point disorder for $H \parallel c$ ({\it e.g.} vortex glass and Bragg glass)~\cite{FisherDS91,Giamarchi95} can be prevented. In the absence of quantum fluctuations, random point disorder can cooperate with the intrinsic pinning effect to stabilize the low-temperature
vortex smectic and vortex solid phases,~\cite{Balents94,Balents95} so that the vortex phase diagram for $H \parallel ab$ would resemble that of the vortex-glass and vortex-liquid phases observed for $H \parallel c$ with a glass transition $H_G (T = 0)$ approaching $H_{c2} (T = 0)$. On the other hand, when field-induced quantum fluctuations are dominant, the vortex phase diagram for $H \parallel ab$ will deviate substantially from predictions solely based on thermal fluctuations and intrinsic pinning, so that strong suppression of the magnetic irreversibility field $H_{irr} ^{ab}$ relative to the upper critical field $H_{c2} ^{ab}$ is expected at $T \to 0$, because the induced persistent current circulating along both the c-axis and the ab-plane can no longer be sustained if field-induced quantum fluctuations become too strong to maintain the c-axis superconducting phase coherence.

Indeed, experimental studies on a wide variety of cuprate superconductors revealed consistent findings with the notion that all cuprate superconductors exhibit significant field-induced quantum fluctuations, as manifested by a characteristic field $H_{irr} ^{ab} (T \to 0) \equiv H^{\ast} \ll H_{c2} ^{ab} (T \to 0)$ and exemplified in Fig.~14(a).~\cite{Beyer07} Moreover, the degree of quantum fluctuations for each cuprate may be expressed in terms of a reduced field $h^{\ast} \equiv H^{\ast}/H_{c2}^{ab}(0)$, with $h^{\ast} \to 0$ indicating strong quantum fluctuations and $h^{\ast} \to 1$ referring to the mean-field limit. Most important, the $h^{\ast}$ values of all cuprates appear to follow a trend on a $h^{\ast} (\alpha)$-vs.-$\alpha$ plot, where $\alpha$ is a material parameter for a given cuprate that reflects its doping level $\delta$, electronic anisotropy $\gamma$, and charge imbalance if the number of CuO$_2$ layers per unit cell $n$ satisfies $n \ge 3$.~\cite{Kotegawa01,Kotegawa01a} Specifically, $\alpha$ is defined by the following:~\cite{Beyer07}
\begin{eqnarray}
\alpha &\equiv \gamma ^{-1} \delta (\delta _o/\delta _i)^{-(n-2)}, \qquad (n \ge 3); \\
\alpha &\equiv \gamma ^{-1} \delta , \qquad \qquad \qquad \qquad (n \le 2) .
\label{eq:alpha}
\end{eqnarray}
In Eq.~(\ref{eq:alpha}) the ratio of charge imbalance in multi-layer cuprates with $n \ge 3$ is given by $(\delta _o/\delta _i)$~\cite{Kotegawa01,Kotegawa01a} between the doping level of the outer layers ($\delta _o$) and that of the inner layer(s) ($\delta _i$). Finally, in the event that $H_{c2} ^{ab} (0)$ exceeds the paramagnetic field
$H_p \equiv \Delta _{\rm SC} (0)/(\sqrt{2} \mu _B)$ for highly anisotropic cuprates, where $\Delta _{\rm SC} (0)$ denotes the superconducting gap at $T = 0$, $h^{\ast}$ is defined by $(H^{\ast}/H_p)$ because $H_p$ becomes the maximum critical field for superconductivity.

Systematic studies of the in-plane irreversibility fields of various cuprate superconductors revealed a universal trend is found for $h^{\ast} (\alpha)$-vs.-$\alpha$, as shown in Fig.~15(a).~\cite{Beyer07} In particular, it is worth noting the $h^{\ast}$-vs.-$\alpha$ dependence in the multi-layered cuprates $\rm HgBa_2Ca_2Cu_3O_x$ (Hg-1223, $T_c$ = 133K), $\rm HgBa_2Ca_3Cu_4O_x$ (Hg-1234, $T_c$ = 125K) and $\rm HgBa_2Ca_4Cu_5O_x$ (Hg-1245, $T_c$ = 108K): While these cuprate superconductors have the highest $T_c$ and $H_{c2}$ values, they also exhibit the smallest $h^{\ast}$ and $\alpha$ values, suggesting maximum quantum fluctuations. These strong quantum fluctuations can be attributed to both their extreme two dimensionality (i.e., large $\gamma$)~\cite{KimMS98,KimMS01} and significant charge imbalance that leads to strong CO in the inner layers.~\cite{Kotegawa01,Kotegawa01a} This notion is corroborated by the muon spin resonance ($\mu$SR) experiments~\cite{Tokiwa03} that revealed increasing antiferromagnetic ordering in the inner layers of the multi-layer cuprates with $n \ge 3$.
%Given that the $\gamma$ values of all Hg-based multi-layer cuprates are comparable,~\cite{Beyer07} the finding of larger quantum fluctuations (i.e. smaller $h^{\ast}$) in Hg-1245 is suggestive of increasing quantum fluctuations with stronger competing order.

\begin{figure}[!]
\centering
\includegraphics[keepaspectratio=1,width=12cm]{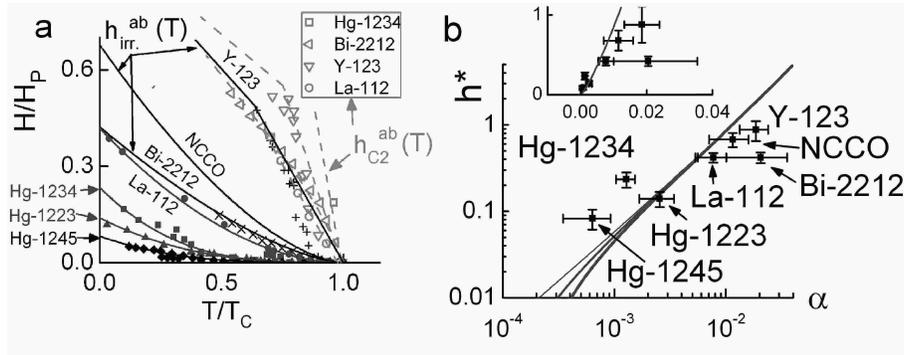}
\caption{(a) Reduced in-plane fields $(H_{irr}^{ab}/H_p)$ and $(H_{c2}^{ab}/H_p)$ vs. $(T/T_c)$ for various cuprates.~\cite{Beyer07} In the $T \to 0$ limit where $H_{irr}^{ab} \to H^{\ast}$, the reduced fields $h^{\ast} \equiv (H^{\ast}/H_p) < 1$ are found for all cuprates Y-123, NCCO, Bi-2212, La-112, Hg-1234, Hg-1223, and Hg-1245 (in descending order).~\cite{Beyer07} (b) $h^{\ast}$ vs. $\alpha$ in logarithmic plot for different cuprates, with decreasing $\alpha$ representing increasing quantum fluctuations. The solid lines are power-law fitting curves given by $5(\alpha - \alpha _c)^{1/2}$, using different $\alpha _c$ = 0, $10^{-4}$ and $2 \times 10^{-4}$ from left to right.}
\end{figure}

Therefore, the investigation of the in-plane magnetic irreversibility in a wide variety of cuprate superconductors reveals strong field-induced quantum fluctuations.~\cite{Beyer07} The macroscopic irreversibility field exhibits dependences on such microscopic material parameters as the doping level, the charge imbalance in multi-layered cuprates, and the electronic anisotropy.~\cite{Beyer07} This finding is consistent with the notion that cuprate superconductors are in close proximity to quantum criticality as the result of coexistence of competing orders and superconductivity in the ground state.

\section{Discussion}

Our consideration so far focuses on the electronic degrees of freedom because they are the dominant contributions in the strongly correlated electronic systems. Nonetheless, it is still important to address possible roles of phonons in the cuprates because of strongly coupled charge, spin and phonon degrees of freedom in these doped Mott insulators.

As discussed earlier, competing orders such as disorder-pinned collective modes play an important role in determining the physical properties of cuprate superconductors. Although it is known to date that the isotope effect in hole-type cuprates anti-correlates with $T_c$ and vanishes near optimal doping,~\cite{ZhaoG01} collective modes such as CDW and PDW that are associated with charge modulations along the Cu-O bonding directions can in fact couple with the longitudinal optical (LO) phonons of the in-plane Cu-O stretching mode in hole-type cuprates, because the latter is known to involve substantial charge transfer from holes in the oxygen 2$p$-orbital to the copper 3$d_{z^2}$-orbital, giving rise to a negative dielectric constant and a local attractive potential.~\cite{Tachiki03} This cooperative effect can result in charge heterogeneity and stabilization of the collective PDW/CDW modes, yielding the pseudogap phenomenon~\cite{Timusk99} above $T_c$ in the under- and optimally doped hole-type cuprates. Further, in hole-type cuprates, the charge-phonon coupling via the LO mode is expected to become more significant in the underdoped limit because of the reduced hole mobility. This conjecture is consistent with the empirical findings of increasing isotope effects with decreasing doping. In contrast, the LO in-plane Cu-O stretching phonon mode cannot incur as significant charge transfer in the electron-type cuprates as in the hole-type, because electron doping in the former primarily resides on the copper site, giving rise to filled 3$d$- and 2$p$-orbitals in the Cu$^+$-O$^{2-}$ bonds, which do not favor phonon-induced charge transfer. We therefore argue that the quantum cooperation effect between the LO phonons and the PDW/CDW collective modes is much weaker in the electron-type cuprates, which is consistent with the absence of zero-field pseudogap in the quasiparticle tunneling spectra of electron-type cuprates. More importantly, given that the LO phonon-induced attractive potential is not universal in the cuprates, and that similar phonon-induced charge transfer is present in other non-superconducting perovskite oxides,~\cite{Tranquada02,Reichardt98} the LO phonons are unlikely solely responsible for the superconducting pairing mechanism in the cuprates, even though they might enhance the $T_c$ values of the hole-type cuprate superconductors.

It is also interesting to address the relevance of competing order scenario to one of the outstanding issues in cuprate superconductors, namely, the anomalous sign reveral in the vortex-state Hall conductivity ($\sigma _{xy}$) for both electron- and hole-type cuprates.~\cite{Iye89,Hagen93,Harris93,Clinton95,Beam98,Beam99} Although quantitative description for the microscopies of the vortex-state Hall conduction remains incomplete, several important
facts have been established. First, the sign reversal is associated with intrinsic physical properties of cuprate
superconductors~\cite{Feigelman95,vanOtterlo95} and is independent of either random~\cite{Vinokur93} or correlated disorder.~\cite{Beam98} Second, the occurrence of sign reversal in $\sigma _{xy}$ has been attributed to the nonuniform spatial distribution of carriers within and far outside the vortex core.~\cite{Feigelman95,vanOtterlo95} Third, the dc vortex-state Hall conductivity of superconducting cuprates is found to be strongly dependent on the doping level, showing anomalous sign reversal in the underdoped regime and no anomaly in the overdoped regime.~\cite{Nagaoka98} This important experimental finding suggests the relevance of the electronic structures within vortices to the Hall conductivity in the superconducting state. Given the STS observation of pseudogap phenomena inside the vortex cores of both hole- and electron-type cuprates~\cite{Beyer09,Teague09} that differs fundamentally from the ``normal core'' approximations in conventional type-II superconductors, it seems natural to attribute the occurrence of sign reveral in the vortex-state $\sigma _{xy}$ to the reduced quasiparticle DOS inside vortices as the result of competing orders. This notion is further corroborated by the increasing sign reveral effect with decreasing doping~\cite{Nagaoka98} because the effects of competing orders become more significant in the underdoped limit.

While the CO scenario can provide a unified phenomenology for the quasiparticle low-energy excitation spectra in both electron- and hole-type cuprates, a number of issues remain unresolved. For instance, the microscopic mechanism for Cooper pairing in the presence of coexisting COs is still unknown. In addition, how different types of COs may emerge in the ground state of different cuprates and how their interplay affects the spectral evolutions with doping level and temperature are not well understood. Moreover, a quantitative account for the quasiparticle LDOS, FT-LDOS and the spectral evolution with increasing magnetic field has yet to be developed. Finally, whether the presence of COs is helpful, irrelevant or harmful to the occurrence of cuprate superconductivity still awaits further investigation.

Although the ultimate answer to the pairing mechanism remains elusive, recent discovery of superconductivity in the iron pnictides with a maximum $T_c \sim 52$ K~\cite{Kamihara08,Takahashi08,Ren08,WenHH08,Sasmal08,Tapp08} may provide useful comparisons for better understanding of the cuprate superconductors. In particular, there are interesting similarities and contrasts between the cuprates and the Fe-based compounds so that parallel studies of the low-energy excitations of both systems are likely to yield useful insights into the fundamental issue of pair formation in superconductors. For instance, the iron pnictides in the stoichiometric limit are magnetic metals rather than insulators,~\cite{Norman08} and they may be doped with either electrons or holes to achieve superconductivity. However, superconductivity in the iron pnictides appears to occur through multi-orbital pairing as opposed to the dominant $d_{x^2-y^2}$ orbital pairing in the cuprates. The discovery of superconductivity in iron chalcogenides~\cite{Hsu08,Mizuguchi08} with Fermi surfaces similar to those of the iron pnictides adds further interest to establishing the pairing symmetry and the role of magnetism in the iron-based superconductors. Indeed, the role of spins in the formation of Cooper pairs~\cite{LeePA06,Scalapino95} has been one of the most debated issues in cuprate superconductors.

\section{Conclusion}

In conclusion, the status of experimental and theoretical investigations of unconventional and non-universal low-energy excitations of cuprate superconductors is reviewed. A phenomenology based on coexisting competing orders with cuprate superconductivity in the ground state appears to provide consistent account for a wide range of experimental findings, and is also compatible with the possibility of pre-formed Cooper pairs and significant phase fluctuations in cuprate superconductors. These findings imply strong interplay between the pairing state and various collective low-energy bosonic excitations in the cuprates, thereby imposing important constraints on the microscopic descriptions for high-temperature superconductivity.

\section*{Acknowledgments}
The work at Caltech is supported by the National Science Foundation through Grants \#DMR-0405088.

\end{document}